\documentclass[11pt,a4paper]{article}
\pdfoutput=1
\usepackage{jheppub}
\usepackage{url}
\usepackage{hyperref}
\usepackage{amsfonts}
\usepackage{epsfig}\usepackage{dcolumn}
\usepackage{bm}
\usepackage{slashed}
\graphicspath{{./figuras/}}
\usepackage{subfigure}

\title{\boldmath Left-Right Symmetry  and Lepton Number Violation at the Large Hadron Electron Collider}

\author{Manfred Lindner,}
\author{Farinaldo S. Queiroz,}
\author{Werner Rodejohann,}
\author{Carlos E. Yaguna}

\affiliation{Max-Planck-Institut f\"ur Kernphysik, Saupfercheckweg 1, 69117 Heidelberg, Germany\\}

\emailAdd{lindner@mpi-hd.mpg.de}
\emailAdd{farinaldo.queiroz@mpi-hd.mpg.de}
\emailAdd{werner.rodejohann@mpi-hd.mpg.de}
\emailAdd{carlos.yaguna@mpi-hd.mpg.de}

\abstract{We show that the proposed Large Hadron electron Collider (LHeC) will provide an opportunity to search for left-right symmetry and establish lepton number violation, complementing current and planned searches based on LHC data and neutrinoless double beta decay. We consider several plausible configurations for the LHeC -- including different electron energies and polarizations, as well as  distinct values for the charge misidentification rate. Within left-right symmetric theories we determine the values of right-handed neutrino and gauge boson masses  that could be tested at the LHeC after one, five and ten years of operation.  Our results indicate that this collider might probe, via the $\Delta L =2$ signal $e^-p\to e^+jjj$, Majorana neutrino masses up to $1$~TeV and  $W_R$ masses up to $\sim 6.5$~TeV.  Interestingly, part of this parameter space is beyond the expected reach of the LHC and of future neutrinoless double beta decay experiments. 
}
    
\begin{document} 
\maketitle
\flushbottom

\section{Introduction}

Left-right (LR) symmetric models are compelling extensions of the  Standard Model (SM) based on the gauge group  $SU(3)_C \otimes SU(2)_L\otimes SU(2)_R\otimes U(1)_{B-L}$ \cite{Pati:1974yy,Mohapatra:1974hk,Mohapatra:1974gc,Senjanovic:1975rk,Senjanovic:1978ev}. Besides naturally appearing in many GUT theories, these models address some of the most pressing problems in particle physics. They give rise to  active neutrino masses via the seesaw mechanism \cite{Minkowski:1977sc,Mohapatra:1979ia,Lazarides:1980nt,Mohapatra:1980yp,Schechter:1980gr}  provide a natural environment for baryogenesis through leptogenesis \cite{Fukugita:1986hr}, incorporate viable dark matter candidates \cite{Berlin:2016eem,Borah:2016uoi,Berlin:2016hqw,Patra:2015vmp,Garcia-Cely:2015quu,Heeck:2015qra}, and directly address parity violation at the weak scale \cite{Senjanovic:1975rk,Senjanovic:1978ev}. The scale at which the left-right symmetry must be  broken is, however, not predicted, implying that the  masses of the new gauge bosons ($Z_R$ and $W_R^\pm$) are arbitrary.  They may in principle lie close to the TeV scale and within the reach of the LHC and of future colliders.

At the LHC, both $Z_R$ and $W_R^\pm$ could be directly produced and searched for \cite{Keung:1983uu,Chiappetta:1993jy,Maiezza:2010ic,Tello:2010am,Nemevsek:2011hz,Esteves:2011gk,Das:2012ii,Chen:2013fna,Arbelaez:2013nga}. In the  minimal left-right symmetric model, the masses of these gauge bosons are related by  $M_{Z^{\prime}} = 1.7 M_{W_R}$ for $g_R=g_L$. Since both gauge bosons share similar production cross sections at the LHC,  the best way to constrain this minimal scenario is by performing $W_R$ searches. These searches have the additional advantage of directly probing  the $SU(2)_R$ breaking scale -- $M_{W_R} = g_R v_R$, where $g_R$ is the $SU(2)_R$ gauge coupling. The main $W_R$ search strategy is based on  dilepton plus dijet data via $qq\to W_R\to \ell N_R\to \ell \ell q\bar q $. So far, no evidence of left-right symmetry has been observed. Using $19.7~\mathrm{fb}^{-1}$ of data at 8 TeV of centre-of-mass energy, the CMS collaboration has excluded $W_R$ masses up to $3$ TeV. The LHC projected limit \cite{Ferrari:2000sp}, for a luminosity of $300~\mathrm{fb}^{-1}$ at $14$ TeV, reaches $W_R$ masses of order $5.6$ TeV.

Further into the future,  an alternative way of searching for LR models is offered by the Larger Hadron electron Collider (LHeC). The LHeC is a project still under design that aims to  combine the intense hadron (proton and ion) beams of the Large Hadron Collider  with a new electron accelerator at CERN. The main advantage of such a setup over the LHC would be the cleaner environment. Currently, the plan for the LHeC features a 60-140 GeV electron beam, with the possibility for electron polarization of up to 80\% \cite{AbelleiraFernandez:2012cc,Bruening:2013bga}, colliding with a 7 TeV proton beam using the LHC tunnel, reaching  $100\, {\rm fb^{-1}}$ integrated luminosity per year.

At the LHeC, the production of Majorana neutrinos via the lepton number violating signal $p\ell^-\to\ell^+jjj$ was investigated in the context of the Type I seesaw mechanism in \cite{Blaksley:2011ey}, within a simplified approach in \cite{Duarte:2014zea}, and in the context of the left-right symmetric in \cite{Mondal:2015zba,Queiroz:2016qmc,Mondal:2016czu}. In the latter, though, the background was not properly considered, producing a very optimistic result.  In this paper, we  aim to assess,  within the left-right symmetric model, the sensitivity reach of the LHeC. To that end, we consider four possible LHeC configurations -- with two different electron energies, $60$ and $140$ GeV, and two values of the polarization fraction, $0\%$ and $80\%$ -- and take into account the relevant background processes in each case. A critical quantity in this regard turns out to be the  charge-flip misidentification rate (MID), which determines the contribution of the main SM background, namely $e^-p\to e^-jjj$. Our results consist of the regions, for each LHeC configuration and for three different values of the MID, in the plane Majorana neutrino mass vs.\ $W_R$ mass for which a signal at the LHeC can be seen at 95\% C.L.\ after one, five, or ten years of data. 

To determine if the LHeC has the potential to probe new regions of the parameter space,  we will compare its expected sensitivity  against current bounds and projected sensitivities at the LHC. We will show that, depending on the configuration and the MID rate achievable, the LHeC may play either a complementary or a leading role in the discovery  of the left-right symmetric model. In the former case, the LHeC could not only confirm the existence of  physics beyond the SM previously discovered at the LHC, but it would also pinpoint its left-right symmetric origin by making use of polarized electrons. In fact, the left-right model predicts that if the initial electron is chosen to be mostly right-handed the signal should get enhanced whereas the SM background must get reduced. {In other words, if a signal is seen at the LHC consistent with a new charged gauge boson, the LHeC could undoubtedly confirm whether or not such signal has a right-handed nature by tuning the polarization. In the latter case,  the LHeC could instead discover new physics not previously seen at the LHC. Our results indicate that this possibility is particularly likely for right-handed neutrino masses around $100-400$ GeV.

The collider signal we investigate is $e^-p\to e^+jjj$, and its observation at the LHeC would not only hint at the left-right symmetric model but also establish lepton number violation. In this regard, one can consider the LHeC and neutrinoless double beta decay experiments as complementary ways to search for $\Delta L=2$ processes. For this reason, we also compare the LHeC sensitivity regions we derived against the current bound from KamLAND-Zen \cite{Asakura:2014lma} and the projected reach of nEXO \cite{Pocar:2015mrz}. We remark here that alternative mechanisms for neutrinoless double beta decay such as the one considered here decouple lifetime predictions for the decay from strong cosmological neutrino mass limits which apply  only to the standard mechanism of light neutrino exchange. 
We will show that the LHeC may also probe lepton number violation beyond the reach of future neutrinoless double beta decay experiments.

Summarizing, the main novelties in our analysis of the left-right symmetric model at the LHeC are the following:
\begin{itemize}
\item[(i)] We discuss several plausible configurations for the LHeC, with and without electron polarization and with different electron energies.
\item[(ii)] We properly take into account the background, pointing out the relevance of the  charge flip MID rate.
\item[(iii)] We outline  the regions of the parameter space for which the LHeC constitutes a complementary or leading search strategy for Majorana neutrinos and the left-right symmetric model.
\end{itemize}

The rest of the paper is organized as follows: In the next section we give a brief overview of the left-right symmetric model. Section \ref{sec:constraints} includes a summary of the most relevant current constraints on this model. Our main results are presented in Section \ref{sec:LHeC} where we display the sensitivity reach for four LHeC configurations and three plausible values of the MID rate, and compare   these regions against those from current and planned experiments. Lastly, we summarize and present our conclusions in section \ref{sec:con}.

\section{Left-right symmetric models} 
\label{sec:LR}

Left-right symmetric theories are extensions of the SM based on the gauge group  
$SU(2)_L \otimes SU(2)_R \otimes U(1)_{B-L}$ that can be realized at the TeV scale.
The electric charge operator is given by 
\begin{equation}
Q = T_{3 L} + T_{3 R} + \frac{B - L}{2}\,,
\label{qlr}
\end{equation}
where $T_{3L/3R}$ are the generators of left-and right-handed isospin. The fermions, which belong to  the fundamental
representation of $SU(2)_{L/R}$, transform as
\begin{eqnarray*}
Q_{L} \sim (3,2,1,1/3),  \hspace{0.2in}     \ell_{L} \sim  (1,2,1,-1) \,,\\
Q_{R} \sim (3,1,2,1/3), \hspace{0.2in}     \ell_{R} \sim  (1,1,2,-1) \,.
\end{eqnarray*}
The usual scalar sector consists of Higgs triplets $\Delta_L$
and $\Delta_R $ as well as of a Higgs bi-doublet $\Phi$, transforming
as \footnote{See \cite{Bambhaniya:2015wna,Bambhaniya:2014cia,Bambhaniya:2013wza,Dev:2016dja,Bambhaniya:2015ipg,Mohapatra:2013cia} for recent limits on the scalar sector of Left-Right models}:  
\begin{equation}
\Phi=
\left(\begin{array}{cc}
\ \phi^0_{11} & \phi^+_{11} \\
\ \phi^-_{12} & \phi^0_{12}
\end{array}\right)
\sim (2,2,0) \,,
\nonumber 
\end{equation}
\begin{equation}
\Delta_L =
\left(\begin{array}{cc}
\ \delta^+_L/\surd 2 & \delta^{++}_L \\
\ \delta^0_L & -\delta^+_L/\surd 2
\end{array}\right) 
\sim (3,1,2)\,,
\end{equation}
\begin{equation}
\Delta_R =
\left(\begin{array}{cc}
\ \delta^+_R/\surd 2 & \delta^{++}_R \\
\ \delta^0_R & -\delta^+_R/\surd 2
\end{array}\right)
\sim (1,3,2) \,.\nonumber
\end{equation} 
Spontaneous symmetry breaking occurs when the neutral components
of those multiplets acquire vacuum expectation values. 
%
The $SU(2)_L \times SU(2)_R \times U(1)_{B-L}$ breaks to $SU(2)_L\times U(1)_Y$ through $\langle\Delta_R \rangle$ followed by $\langle \Phi \rangle$ which yields $U(1)_{\rm em}$. Moreover, $\Delta_L$ acquires a small vev, $\langle\Delta_L \rangle \propto v^2/v_R$, that contributes  to active neutrino masses.  The SM interactions are reproduced with $M_W^2 = g_L^2 v^2 \equiv g_L^2 (v_1^2 + v_2^2)$, $M_Z=M_W/C_W$, where $v_1$ and $v_2$ are the vevs of the neutral scalars in the bidoublet and $g_L=e/S_W$ is the $SU(2)_L$ gauge coupling, with $C_W \, (S_W)$ being the cosine (sine) of the Weinberg angle. As for the right-handed gauge bosons, their masses are found to be
\begin{equation}
 M_{W_R} =  g_R \, v_R \,,
\end{equation}
 \begin{equation}
 M_{Z_R} = \frac{\sqrt{2} g_R/g_L \, \, M_{W_R}}{\sqrt{(g_R/g_L)^2 -
     \tan \theta_W^2}}\,.
 \end{equation}
In addition, a discrete left-right symmetry, e.g.\ parity, is often assumed so that $g_L = g_R \equiv g$, with the result that $M_{Z_R}\simeq 1.7 M_{W_R}$ (it is possible however to construct models with $M_{Z_R}\ll M_{W_R}$, see Ref.\ \cite{Patra:2015bga} and references therein). Since the $W_R$ is, in this minimal framework,  the lighest new gauge boson and its mass  is directly connected to the left-right symmetry breaking scale, $W_R$ search strategies are typically the most efficient way to constrain left-right theories. 
   
For most phenomenological analysis, the key parts of the Lagrangian are the charged current interactions, 
\begin{align}
\mathcal L_{W}  = & \frac{g}{\sqrt{2}} \left(
\bar l_L U_{L}^\dag  \slashed{W}\!_L l'_L +
\bar l_R  U_{R}^\dag \slashed{W}\!_R l'_R\right)
+\text{h.c.}\, +\nonumber\\
&\frac{g}{\sqrt{2}} \left(
\bar Q_L V_{L}^\dag  \slashed{W}\!_L Q'_L +
\bar Q_R  V_{R}^\dag \slashed{W}\!_R Q'_R\right)
+\text{h.c.},
\label{eq:lagrangian}
\end{align}
where $U_{L/R}$ is the PMNS mixing matrix for the left-handed and
right-handed leptons,  
whereas $V_{L/R}$ is the CKM 
matrix for the left and right-handed quarks, respectively.

Notice that we assume no mixing between the SM $W$ and the $W_R$
gauge bosons.   On the one hand, stringent limits on such a mixing, from a multitude of data \cite{Deppisch:2015qwa}, already exist.

\section{Current Constraints}
\label{sec:constraints}

Several constraints, from collider and precision data, already exists on left-right models. 
Here we will briefly review the most stringent limits related to our analysis. 

\subsection{$W_R$ searches at the LHC}
At present, the most stringent collider limits on left-right models stem from searches at the LHC \cite{Aad:2014cka,Khachatryan:2014dka}. These searches rely on $W_R$ resonances decaying into right-handed neutrinos $N_R$ and charged leptons, and implicitly assume  that $M_{W_R} >M_{N_R}$. The signal consists of dileptons plus dijets through the process  $pp \rightarrow W_R \rightarrow l N_l\rightarrow ll  W_R^{\ast} \rightarrow l l \bar{q}\bar{q}$. Final states with two isolated leptons and at least two jets in the final state are subject to a sizeable SM background arising from $pp \rightarrow WW +{\rm jets},ZZ+{\rm jets},WZ +{\rm jets},\bar{t}t$. The leptons might come from the $W$ or $Z$ decay, or simply from the $\tau$ decay.  Moreover, since the $W_R$ will be resonantly produced, the two leptons in the final state come from the decay of heavy particles and hence their invariant mass is not peaked. An efficient way to reduce SM background without losing much of the signal is to apply  hard cuts on the transverse energy of the jets and to enforce the invariant mass of the leptons to be sufficiently large.

Using an integrated luminosity of $\mathcal{L}=4.7$ fb$^{-1}$ and 7 TeV of centre-of-mass energy, the ATLAS collaboration ruled out $W_R$ masses up to 2.5 TeV at 95\% C.L.\ \cite{Aad:2014cka}. Similarly, CMS using much more  data, $\mathcal{L}=19.7$ fb$^{-1}$ with 8 TeV of centre-of-mass energy  excluded $W_R$ masses up to 3 TeV at 95\% C.L.\ \cite{Khachatryan:2014dka}. In the figures presented in the next section,  the parameter space excluded by the  CMS 8 TeV $eejj$ data is displayed  as a dark green region. Notice that the  LHC bounds tend to be  quite weak when the Majorana neutrinos are light ($M_{N_R} < 100$ GeV or so), offering  a discovery window  for the LHeC, as we will show in the next section. The reason for this behavior is that as the mass of the Majorana neutrinos is reduced, the transverse energy of the jets might decrease to levels at which the signal events do not pass the cuts imposed by the collaboration to reduce the SM background. In other words, the signal has small acceptance for such light Majorana neutrinos. 

Moreover, the quoted results from CMS do not enforce same-sign charge assignments \cite{Khachatryan:2014dka}. The signal is simply determined by the total number of events $l^{\pm}l^{\pm} jj$, which is then compared to SM background expectations in order to draw exclusion regions in the Majorana neutrino vs.\ $M_{W_R}$ mass plane. We should remark that ATLAS has actually performed a search for Majorana neutrinos, focused on same sign-dilepton resonances \cite{Aad:2015xaa}. Nevertheless for right-handed neutrinos below 1 TeV the obtained limits are weaker for the $eejj$ channel but slightly stronger for the $\mu\mu jj$ one. By enforcing same-sign dileptons one can significantly reduce the SM background, but due to worse acceptance/efficiency for dielectron channels the reach for $eejj$ ends up being similar to the one without charge requirements. This slight improvement in the  $\mu\mu jj$ is due to the relative better acceptance/efficiency caused by the muon silicon tracker. Hereafter, we will use the limits obtained from the $eejj$ studies above, and compare with our finding for the $e^+jjj$ signal at the LHeC.

Besides the already mentioned  CMS limits,  the LHC projected limit for $300$ fb$^{-1}$ and 14 TeV centre-of-mass energy derived in \cite{Ferrari:2000sp} will also be displayed in our figures. According to it,  the LHC has the potential to discover a new charged gauge boson with mass up to $\sim 5.6$~TeV.  It is important to stress, though, that even if the LHC were to observe a signal consistent with a $W_R$ gauge boson, it would be rather hard to establish a possible right-handed nature. At the  LHeC, this task would be much easier thanks to the possibility of  using  polarized electron beams, which would increase (decrease) the signal over background ratio for right-handed (left-handed) polarized electrons. Thus, even if a $W_R$ signal were observed at the LHC, the LHeC could still play an important complementary role in establishing its possible left-right symmetric origin.    

For completeness we point out that similar analysis in the context of lepton number violation have been performed for the HERA and LEP experiments in \cite{Buchmuller:1991tu,Ingelman:1993ve,Buchmuller:1990vh,Buchmuller:1992wm,Flanz:1999ah}, and others in the context of Left-Right models \cite{Helo:2013ika,Maiezza:2014ala,Fowlie:2014mza,Dutta:2014dba,Parida:2014dla,Helo:2015ffa,Gluza:2015goa,Brehmer:2015cia,Deppisch:2015cua,Patra:2015bga,Gluza:2016qqv,Lindner:2016lpp}.



\subsection{Neutrinoless Double Beta Decay}
\begin{figure}[t!]
\center
\includegraphics[width=0.5\columnwidth]{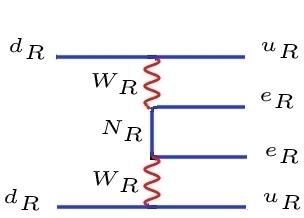}
\caption{Feynman diagram that accounts for the purely right-handed contribution to $0\nu \beta \beta$ in the left-right symmetric model.}
\label{feynfig1}
\end{figure}
Neutrinoless double beta decay experiments are sensitive to neutrino masses and to lepton number violation through the decay mode $(A,Z) \rightarrow (A,Z + 2) +2e^-$, as has been  extensively reviewed in the literature \cite{Vergados:2002pv,Simkovic:2007vu,Avignone:2007fu,Rodejohann:2011mu,Elliott:2012sp,Bilenky:2012qi,Vergados:2012xy,Rodejohann:2012xd,Deppisch:2012nb,Vogel:2012ja,Schwingenheuer:2012zs,Petcov:2013poa,Cremonesi:2013vla,Pas:2015eia}.  Currently, this search strategy is  at a very promising state due to the various operating and planned experiments. The old limit on this decay rate, which was set by the Heidelberg-Moscow experiment in 2001 \cite{KlapdorKleingrothaus:2000sn}, has been continuously improved  \cite{Gando:2012zm,Auger:2012ar,Agostini:2013mzu,Albert:2014awa} and future projects will substantially raise existing limits on the neutrinoless double beta decay lifetime.

Within the left-right symmetric model, many analyses have already  emphasized the role of these experiments in constraining the allowed parameter space \cite{Dev:2013vxa,Dev:2014xea,Deppisch:2015qwa,Dev:2015vra,Awasthi:2015ota,Nemevsek:2011aa,Tello:2010am,Peng:2015haa,Han:2012vk,Helo:2013dla,Teixeira:2014jza,Abada:2013bpa,Abada:2012re,Abada:2008gs,Das:2012ii,Bambhaniya:2015ipg}. Among the several diagrams that may contribute to neutrinoless double beta decay ($0\nu\beta\beta$) in left-right symmetric models (see \cite{Barry:2013xxa} for a recent detailed study), we will focus on the purely right-handed Majorana neutrino contribution shown in Fig.\ \ref{feynfig1}, which is entirely determined by the right-handed gauge interactions (the other contributions related to the left-right mixing and scalar triplets  can be easily suppressed and will be neglected). In this setup, the non-observation of this lepton number violating process can be used to set bounds on the masses of the $W_R$ and the Majorana neutrino. Using the current limit from Kamland-Zen of $2.6\times 10^{25}$ yrs \cite{Asakura:2014lma} for the decay of $^{136}$Xe,  and the projected limit from  nEXO of $6\times 10^{27}$ yrs \cite{Pocar:2015mrz} one can find the current (projected) bound of \cite{Barry:2013xxa,Dev:2013vxa}
\begin{equation}
G_F^2 M_W^4 \frac{\left| V_{ei}^2 \right| }{M_{N_{Ri}} M_{W_R}^4} \leq 
1.4 \times 10^{-16} \,\,(9.5 \times 10^{-18})\,\, {\rm GeV^{-5}}\,.
\end{equation}
Hereafter we will take $V_{e1}=1$ in order to compare our findings concerning $0\nu\beta\beta$ decay with collider bounds on equal footing.  LHC searches for left-right symmetry, in fact, neglect the left-right mixing and assume $100\%$ branching ratio into the lightest heavy neutrino, which is equivalent to taking $V_{e1}=1$. In the figures presented in the next section, the region excluded by KamLAND-Zen is shown pink-shaded  whereas the projected limit from nExo is displayed as a dashed pink line.


\subsection{Meson Mixings}

In the commonly adopted case of left-right parity the lower limit on the mass of the right-handed gauge boson from the $K$ mass difference \cite{Maiezza:2010ic} is about 3 TeV \cite{Maiezza:2014ala}, assuming equal mixing matrices for left- and right-handed quarks. 
The precise limit depends somewhat on the choice of the discrete left-right symmetry (parity or charge conjugation). These discrete symmetries play a role in determining the relation between the Yukawa couplings of the theory, thus restricting the left- and right-handed quark and lepton mixing matrices, which set the aforementioned bound. We will exhibit this limit in the upcoming plots as a vertical dashed red line. 

\section{LHeC Prospects}
\label{sec:LHeC}

The Large Hadron Electron Collider \cite{AbelleiraFernandez:2012cc}  (LHeC) is a proposed  electron-proton collider with an electron beam of 60 GeV to possibly 140 GeV energy colliding with a proton beam of 7 TeV from the LHC. Interestingly, the LHeC is projected to surpass the integrated luminosity of the former $ep$ collider HERA  by two orders of magnitude, reaching  $\mathcal{L}=100$ fb$^{-1}$ per year. In what follows we will investigate the LHeC sensitivity to the left-right symmetric model for $\mathcal{L}=100$ fb$^{-1}$, $\mathcal{L}=500$ fb$^{-1}$ and $\mathcal{L}=1$ ab$^{-1}$, corresponding respectively to one, five and ten years of LHeC operation.

\begin{figure}[t]
\center
\includegraphics[width=0.5\columnwidth]{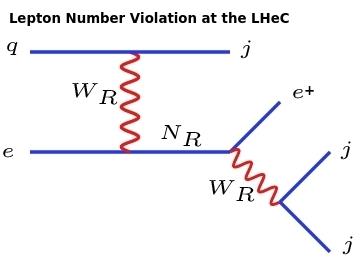}
\caption{Feynman diagram for the signal $e p \rightarrow e^+  jjj$ at the LHeC. Since LHeC will be designed to allow right-handed polarized electron beams, it constitutes an exciting opportunity to probe lepton number violation and heavy Majorana neutrinos.}
\label{feynfig}
\end{figure}

The relevant signal, which violates lepton number by two units\footnote{Similar processes at HERA were discussed e.g.\ in \cite{Buchmuller:1991tu,Flanz:1999ah}.}, is illustrated in Fig.\ \ref{feynfig}. A $W_R$ is exchanged between the initial particles resulting in the production of an on-shell Majorana neutrino, which subsequently decays into $e^+ jj$. Similarly to the LHC analyses, we do not include the decays of the Majorana neutrinos into other charged leptons ($\mu$ and $\tau$). In this way,  the number of free parameters remains small and the comparison with other experiments is transparent.  Since the centre-of-mass energy at the LHeC is of order $1.3$ TeV, Majorana neutrinos with  masses up to $\sim 1$~TeV can be in principle probed. The $W_R$ gauge boson, on the other hand, is not produced on shell -- unlike at the LHC -- so its effects could be observed at the LHeC even if its mass is much larger than the centre-of-mass energy.

A key aspect concerning the LHeC is the possibility of using polarized electrons, but it is not yet clear what their maximum polarization, if any, will eventually be. The proposal\footnote{See \url{http://lhec.web.cern.ch/talks-seminars} for recent talks on LHeC prospects.} described in \cite{AbelleiraFernandez:2012cc} discusses polarizations of $P_e=60\%$ and $P_e=80\%$. Within the left-right model, using a polarized electron beam would certainly be an advantage because one could choose the initial electron to be mostly right-handed, enhancing the signal (which is induced by a right-handed current) and at the same time reducing the SM background. In addition, the polarized electron would offer a unique opportunity to unveil the left-right symmetric origin of an observed signal \cite{Mondal:2015zba}.  To be as general as possible, in our analysis we will consider two different polarizations: $0\%$ (non-polarized) and $80\%$. The result for $60\%$ polarization will be just slightly weaker than  that for $80\%$.

In all, we will investigate  four different configurations for the LHeC, correspoding to two possible  electron energies and two different polarizations:

\begin{itemize}
\item {\bf Configuration 1}: 60 GeV electron beam non-polarized;
\item {\bf Configuration 2}: 60 GeV electron beam 80\% polarized;
\item {\bf Configuration 3}: 140 GeV electron beam non-polarized;
\item {\bf Configuration 4}: 140 GeV electron beam 80\% polarized
\end{itemize}

Without doing any calculations we can assert that configuration 1 is  the least promising whereas configuration 4 offers the best prospects to probe the left-right model. On the other hand, whether configuration 2 is or not more suitable than configuration 3 cannot be ascertained beforehand and requires a detailed analysis, as the one presented in this section. One of our goals is precisely to settle this question. In addition, we want to establish the LHeC reach for each of these configurations and, based on that, to find out what the LHeC role might be in the search for the left-right symmetric model.

To do so, it is of utmost importance to properly treat the possible backgrounds at the LHeC. Since the signal $e^-p\to e^+ jjj$ violates lepton number, it has no theoretical backgrounds from SM processes. That does not mean, however, that the search is background-free.  Due to detector effects,  this signal still suffers from SM backgrounds stemming from electron charge misidentification (MID or charge flip), where a $e^-$ is misidentified as an $e^+$ --this crucial point was overlooked in \cite{Mondal:2015zba}. This effect results in a significant background from processes of the type $e^-p\to e^-jjj$, which do not violate lepton number. The charge misidentification rate at the LHeC plays, therefore, a crucial role in determining its sensitivity to the left-right symmetric model -- it needs to be small enough to suppress the background. After sifting through the LHeC proposal and contacting several members of the LHeC collaboration, we realized that is unclear up to this point what the charge flip MID rate will be at the LHeC. Albeit, it is expected to reach levels below those found at the LHC, due to the the cleaner environment and the lower centre-of-mass energy. At the LHC, as a result of the high $p_T$ electrons, the MID rate is sizable, of order $1\%$ \cite{Khachatryan:2015hwa}. In the absence of more precise data or estimates in this direction, we  work under three MID scenarios for the LHeC: a conservative one with a MID rate of $1\%$, the same as at the LHC; a realistic one with a MID rate of $0.1\%$, which seems plausible for  the LHeC; and an optimistic one with a MID rate of $0.01\%$, which may be difficult to achieve. 


An additional possible background arises from the jet fake rate, but it was found to be, for the selection cuts discussed below, much smaller than  the MID one  and will therefore be ignored in the following.  As for the SM background $e^+jjj\nu_e \nu_e$ studied in \cite{Blaksley:2011ey,Duarte:2014zea}, we checked that it is suppressed in our treatment, due to the hard jet $p_T$ cut we impose.



The signal $(ep \rightarrow e^+ jjj)$ and background $(ep \rightarrow e^- jjj)$ were both  simulated at parton level using Calchep \cite{Belyaev:2012qa}. In the calculation of the SM background, we accounted for collinear divergences using the Weizs\"acker-Williams approximation \cite{Kniehl:1990iv}, which consist of replacing the electron with a photon with corresponding momentum and parton distribution function, as implemented in Calchep \cite{Belyaev:2012qa}.  The resulting signal and background  were then fed into Pythia for clustering and hadronization \cite{Sjostrand:2014zea}, assuming a flat 100\% acceptance times efficiency for the signal. We require the lepton jets to be isolated within a cone $\Delta R^{jj} > 0.5$ and $\Delta R^{je} > 0.5$. After several attempts, we found that the following cuts provide the best signal/noise ratio for the four LHeC configurations: 
\begin{equation}
p_T^j > 50\, {\rm GeV},\quad p_T^e > 10 \, {\rm GeV},\quad \Delta R^{jj} > 0.5,\quad \Delta R^{je} > 0.5,\quad |\eta^j| < 2.5, \quad |\eta^e| < 2.5.
\end{equation}
Defining the statistical significance as $S/\sqrt{S+B}$, where $S$ and $B$ are the number of signal and background events, we can then draw -- for a given LHeC configuration, luminosity, and MID rate -- the region  in the $M_{N_R}$ vs.\ $M_{W_R}$ plane leading to the observation of a lepton number violation signal at 95\% C.L.\ with at least one event. The 95\% C.L.\ signal regions that yield less than one event for the luminosities we considered were ignored.

In the following subsections, where our main results are presented, we determine and display these regions for the four different LHeC configurations. For each given configuration, we include three panels corresponding to specific values of the MID rate. In each of those panels, the sensitivity regions are displayed for one, five and ten years of data.

\subsection{Configuration 1: $E_e =60\,\mathrm{GeV}$, $P_e= 0$}
This  configuration refers to a $60$~GeV unpolarized electron beam and is the most pessimistic among the four we consider. Thus, it can give us a good idea of the minimum reach that can be achieved at the LHeC. 

Figure \ref{60GeVPe00} shows the 95\% C.L.\ regions delimiting the LHeC sensitivity to the lepton number violating signal $ep \rightarrow e^+ jjj$ mediated by a $t$-channel $W_R$ -- see Fig.\ \ref{feynfig}.  The three panels display  these regions for different MID rates: 1\% (left panel), 0.1\% (right panel) and 0.01\% (bottom panel). In each panel,  the signal regions are obtained for three different luminosities,  corresponding to one ($100$ fb$^{-1}$), five ($500$ fb$^{-1}$) and ten years ($1$ ab$^{-1}$) of LHeC operation. The shape of the signal regions are determined by a combination of several factors including: (i) parton distribution functions, which favor Majorana neutrino masses around hundred of GeV's; (ii) the dependence on the $W_R$ mass; (iii) the phase space integral; (iv) the centre-of-mass energy, $\sim 1.3$ TeV.

For comparison, we have overlaid the LHeC reach with the current (projected) limits previously discussed, stemming from  KamLAND-Zen (nEXO) for neutrinoless double beta decay experiments, and from LHC8TeV (LHC14TeV) for collider searches.

\begin{figure*}[t]
\includegraphics[width=0.5\columnwidth]{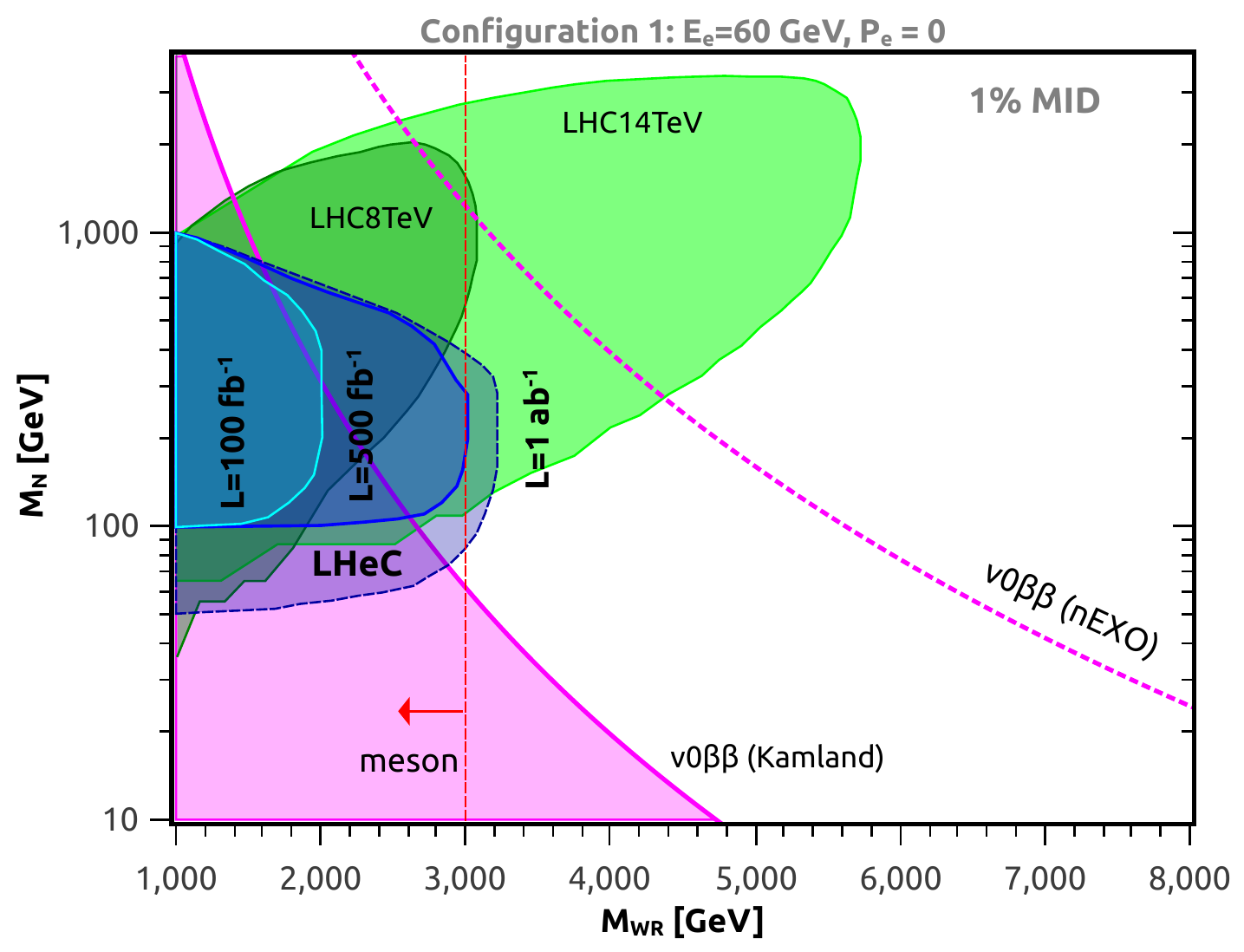}
\includegraphics[width=0.5\columnwidth]{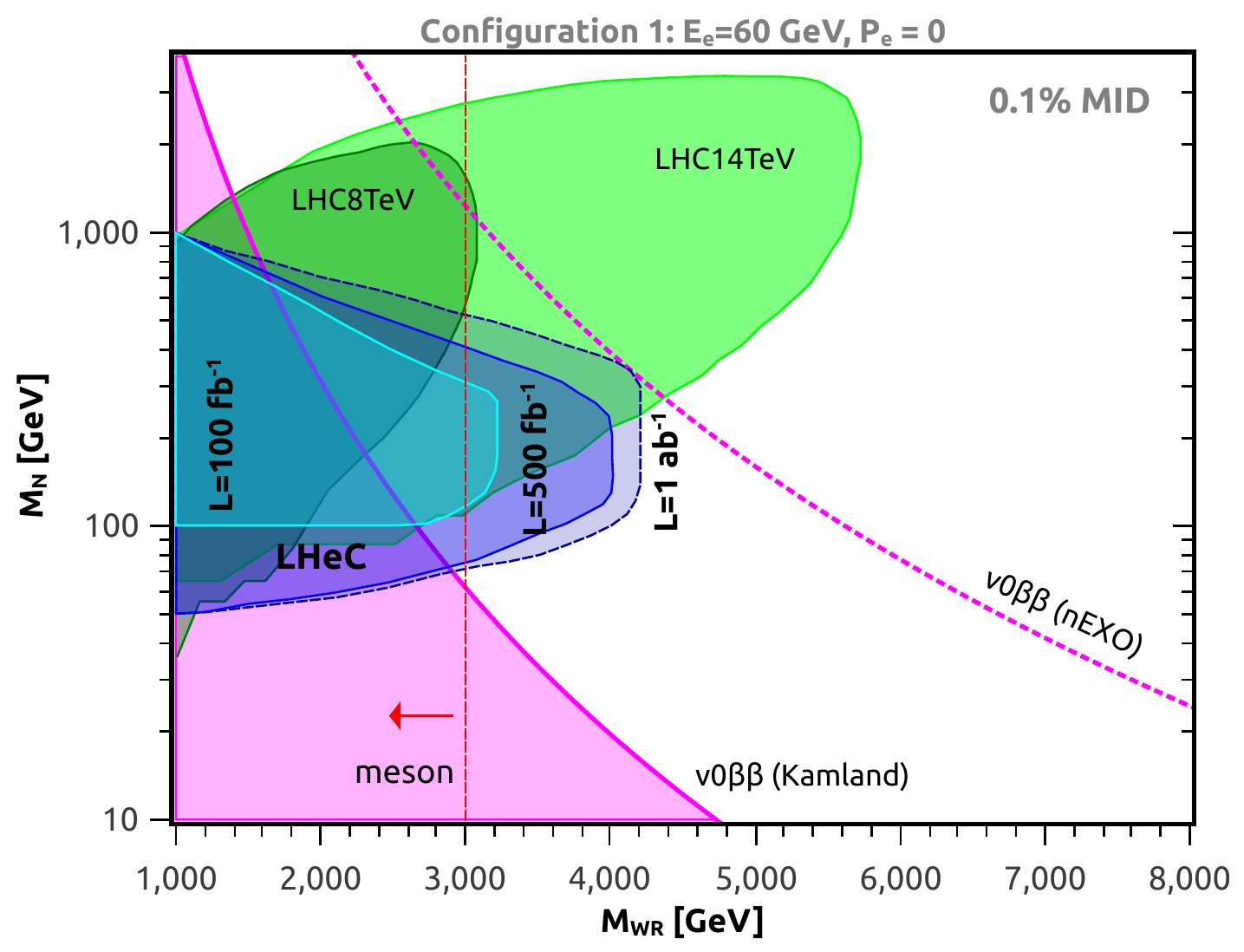}
\center
\includegraphics[width=0.5\columnwidth]{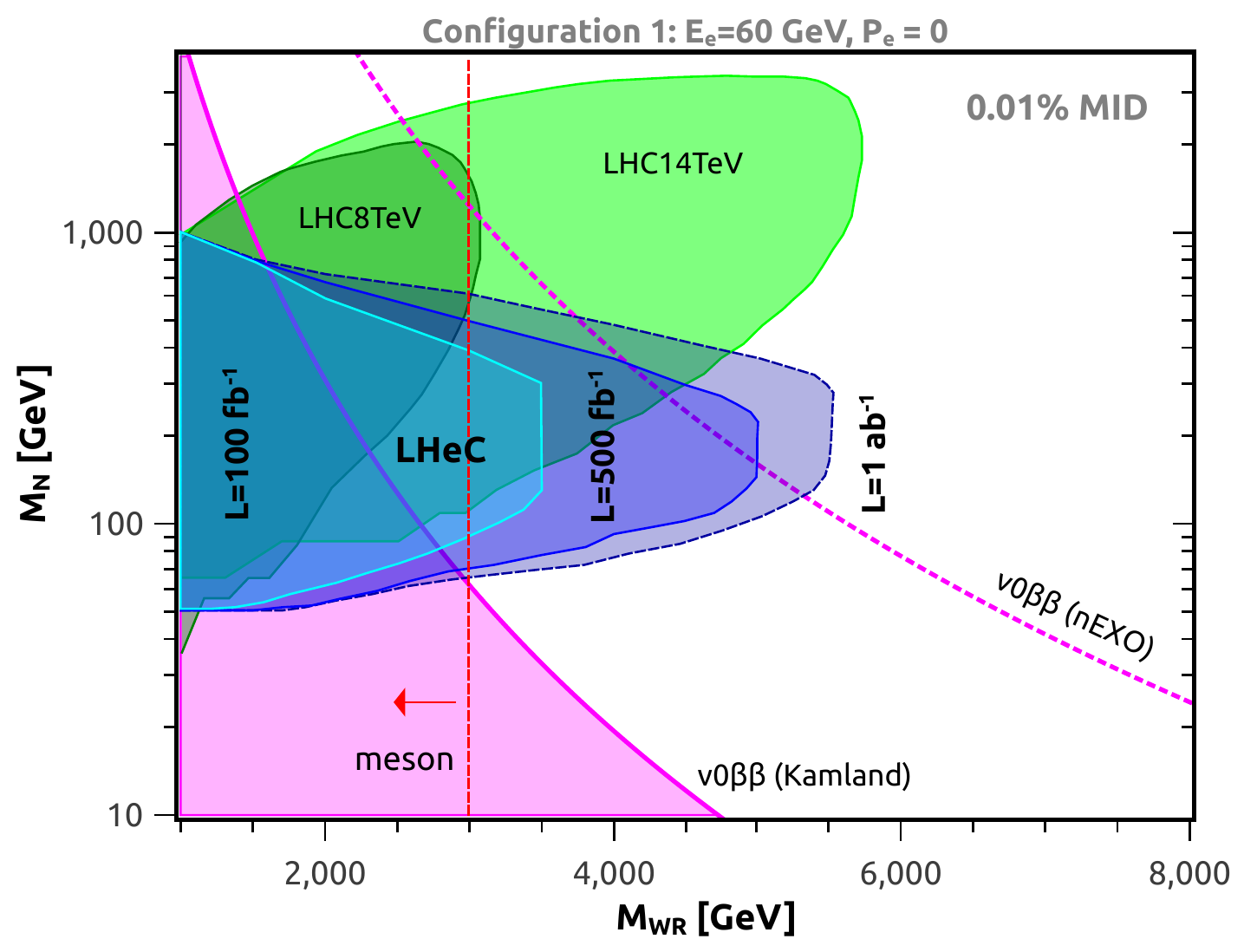}
\caption{{\bf Configuration 1}: 95\% C.L.\ region probed by LHeC for $\mathcal{L}=100$ fb$^{-1}$ (light blue), $500$ fb$^{-1}$ (blue), $1000$ fb$^{-1}$ (dark blue). {\it Left:} assumes $1\%$ MID rate; {\it Right:} adopts a 0.1\% MID rate; {\it Bottom:} uses a 0.01\% MID rate. We have superimposed current limits from $eejj$ (dark green), KamLAND-Zen measurement (shaded pink); and projected limits from: $eejj$ LHC data (light green) and nEXO (dashed pink line) sensitivity.}
\label{60GeVPe00}
\end{figure*}

From the left panel of Fig.\ \ref{60GeVPe00} it can be read  that the LHeC may probe $W_R$ masses up to $2$~TeV, $3$~TeV and $3.2$~TeV respectively for $\mathcal{L}=100$ fb$^{-1}$ (light blue), $500$ fb$^{-1}$ (blue), and  $1000$ fb$^{-1}$ (dark blue).  A large part of these regions have already been excluded by existing searches, and those that have not entirely lie within the expected sensitivity of future experiments, particularly of nEXO. In this rather pessimistic scenario, with no polarization and a MID rate of $1\%$, the LHeC would  play a complementary role to other searches, and may help identify the left-right origin of the new physics signals that would have been already observed at the LHC and in nEXO.   If the MID rate is instead $0.1\%$ (right panel), the situation improves quite a bit, with the LHeC reach extending to  $W_R$ masses of up to 4.2 TeV after  $10$ years of data. In this case, we already find a non-negligible region, for $M_N\sim 100$ GeV,  where the LHeC can improve over the expected reach of the LHC. Even in this case, though, the entire region lies within the projected sensitivity of nEXO. To find, within this configuration, regions where the LHeC could perform better than nEXO we need to reduce further the MID rate and  use at least five years of data, as illustrated in the bottom panel.   


Thus, in order to ensure that the LHeC probes  regions of the parameter space beyond  the reach of existing and planned experiments, the MID rate within this configuration has to be kept below $0.1\%$. As we now show, the other configurations, with higher energy or polarized electrons, offer better prospects.

\subsection{Configuration 2: $E_e =60 {\rm \, GeV}$, $P_e= 80\%$}

\begin{figure*}[t]
\includegraphics[width=0.5\columnwidth]{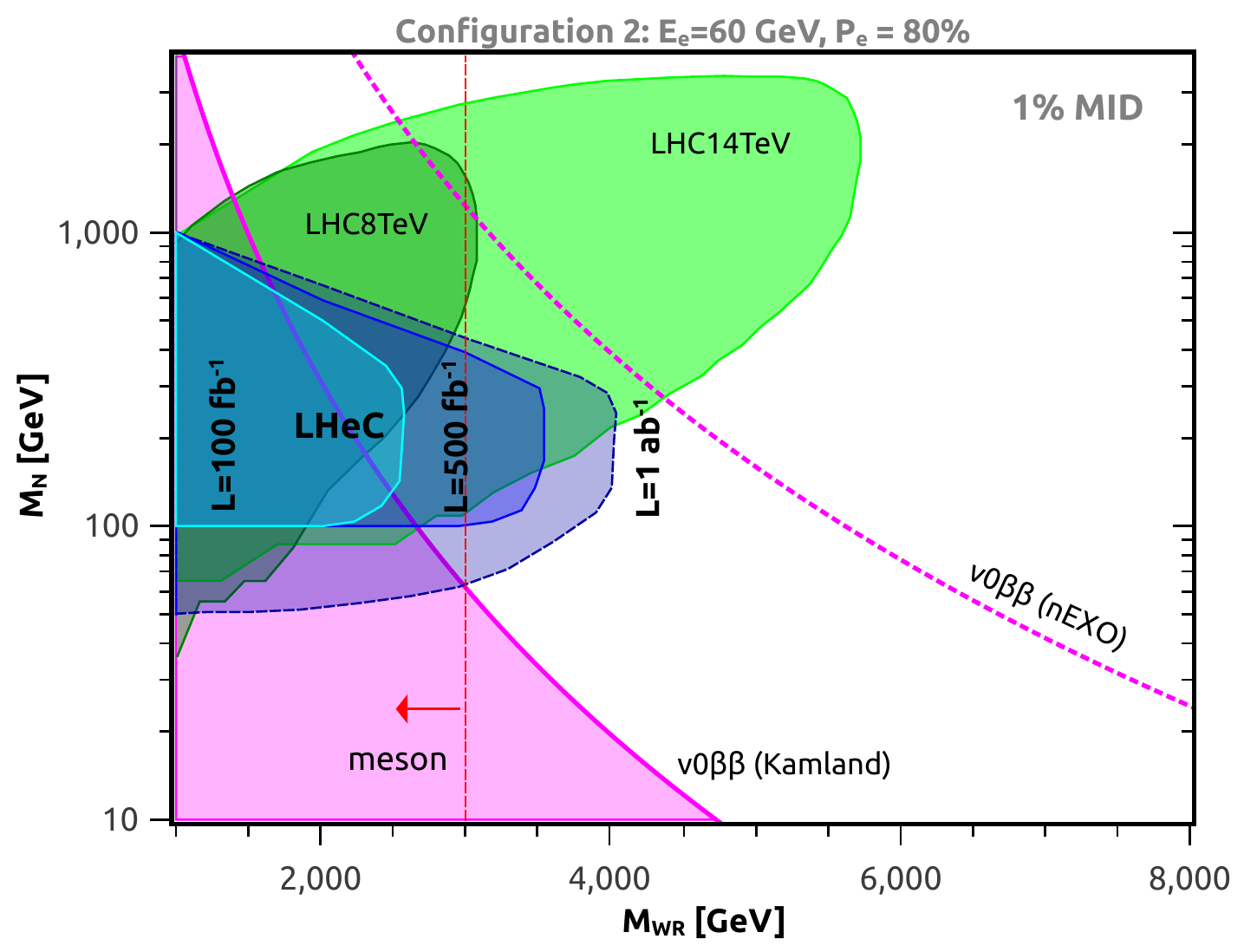}
\includegraphics[width=0.5\columnwidth]{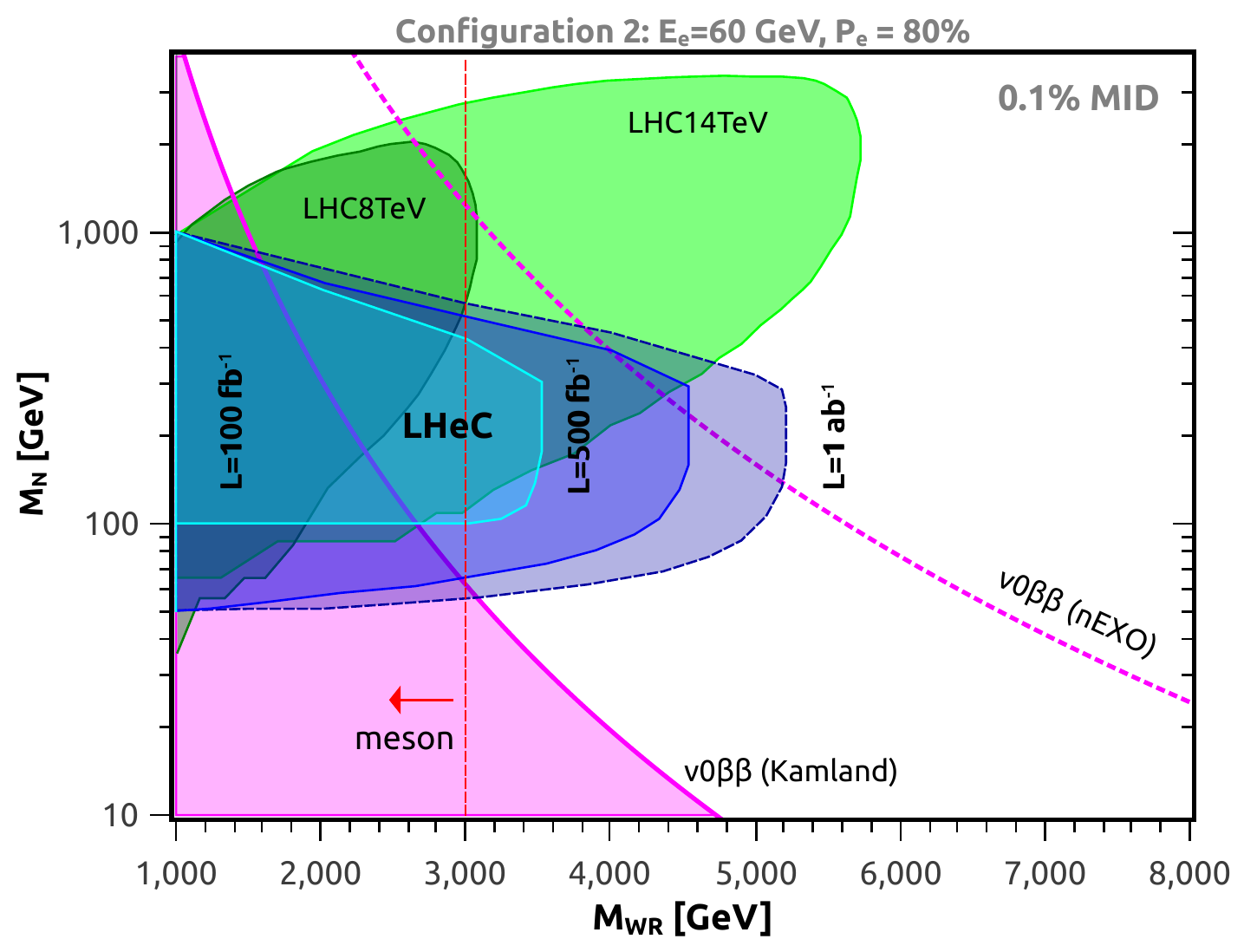}
\center
\includegraphics[width=0.5\columnwidth]{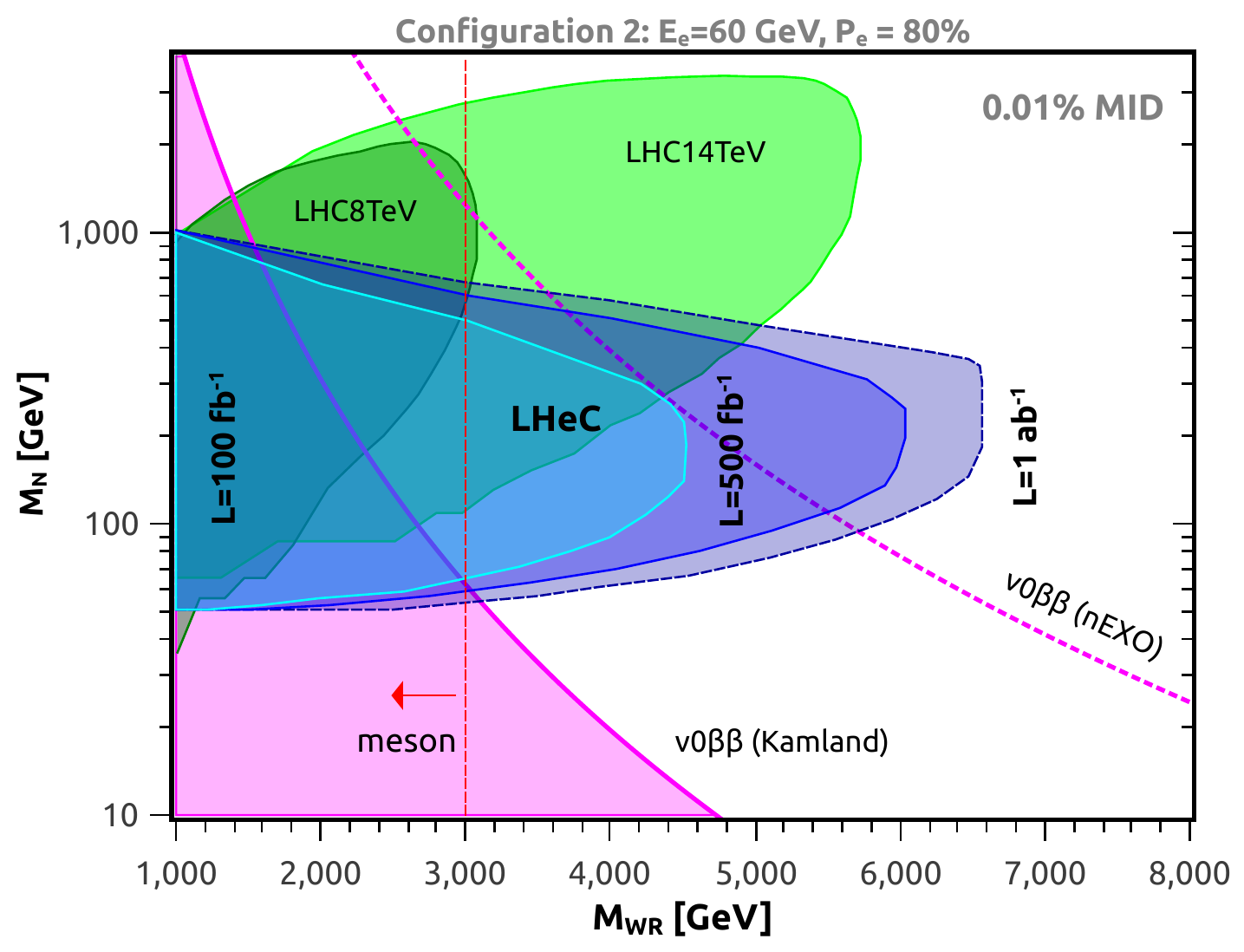}
\caption{{\bf Configuration 2}: 95\% C.L.\ region probed by LHeC for $\mathcal{L}=100$ fb$^{-1}$ {(light blue)}, $500$ fb$^{-1}$  (blue), 
$1000$ fb$^{-1}$ (dark blue). {\it Left:} assumes $1\%$ MID rate; {\it Right:} adopts a 0.1\% MID rate; {\it Bottom:} uses a 0.01\% MID rate. We have superimposed current limits from $eejj$ (dark green), KamLAND-Zen measurement (shaded pink); and projected limits from: $eejj$ LHC data (light green) and nEXO (dashed pink line) sensitivity.}
\label{60GeVPe04}
\end{figure*}

In this configuration, the electron energy is the same but the beam is assumed to be $80\%$ polarized.  Ramping up the polarization to $80\%$ increases the signal cross section by a factor of two or so, while reducing the SM background. We emphasize though that the most important aspect of increasing the electron polarization is not necessarily the improvement on the LHeC reach but the fact that by tuning the polarization one can decisively determine whether or not a possible signal increases (decreases) as we use right-handed (left-handed) polarized electron. In this way one can establish whether the signal has a left-right origin. 

By comparing the sensitivity regions with those obtained for configuration 1, we can get a good idea of how important the polarizations effects are regarding the reach to $W_R$ masses. These regions are shown  in Fig.\ \ref{60GeVPe04} using the same conventions as before. After 10 years of data, the maximum $W_R$ mass that can be probed at the LHeC has increased to about $4$, $5.5$ and $6.5$ TeV respectively for a MID rate of $1\%$, $0.1\%$ and $0.01\%$. Interestingly, we see that in this case a MID rate of $0.1\%$ (right panel) is enough to ensure that new regions of the parameter space can be explored at the LHeC. Such regions feature Majorana masses of order $200$ GeV and $W_R$ masses between $4.5$ and $5.5$ TeV.  Furthermore, for a MID rate of $0.01\%$ (bottom panel), the LHeC could probe $W_R$ masses significantly beyond the expected reach of the LHC. 

As our results illustrate, using a polarized electron beam at the LHeC would certainly increase the probability of producing Majorana neutrinos and establishing lepton number violation within the left-right symmetric model. Still, the MID rate is critical and needs to be small enough to guarantee that new regions of the parameter space are tested. 


\subsection{Configuration 3: $E_e =140 {\rm \,GeV}$, $P_e= 0$}
\begin{figure*}[!t]
\includegraphics[width=0.5\columnwidth]{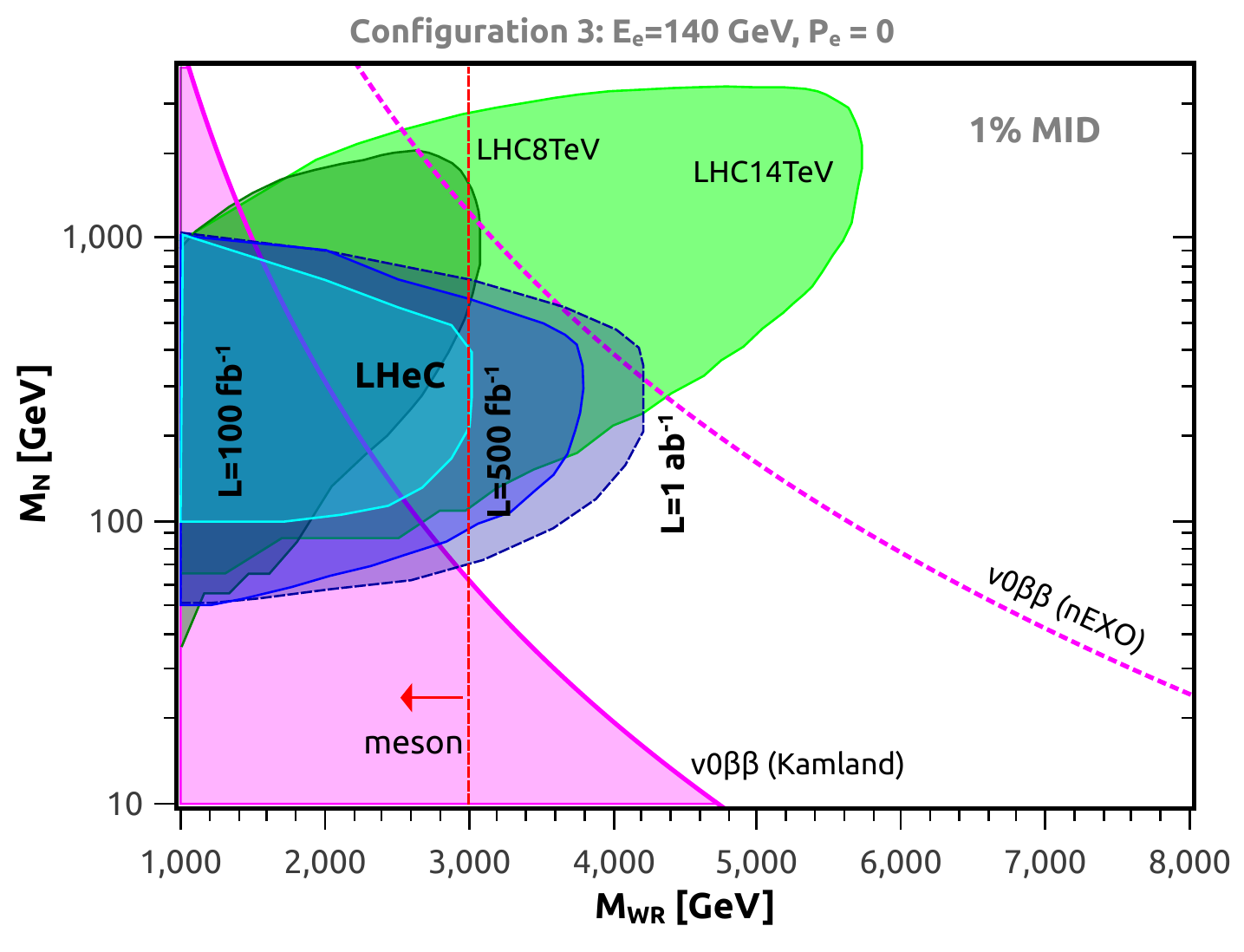}
\includegraphics[width=0.5\columnwidth]{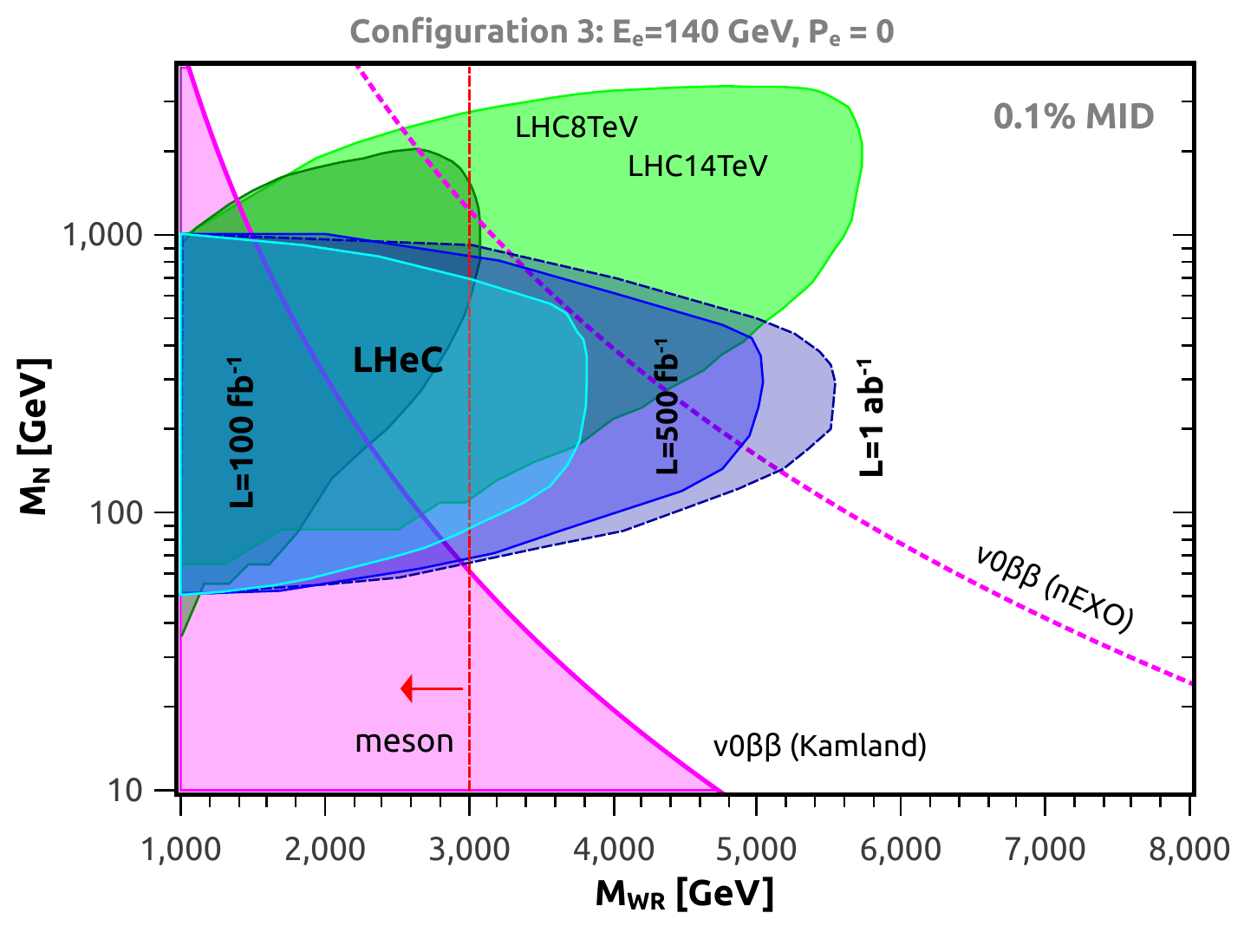}
\center
\includegraphics[width=0.5\columnwidth]{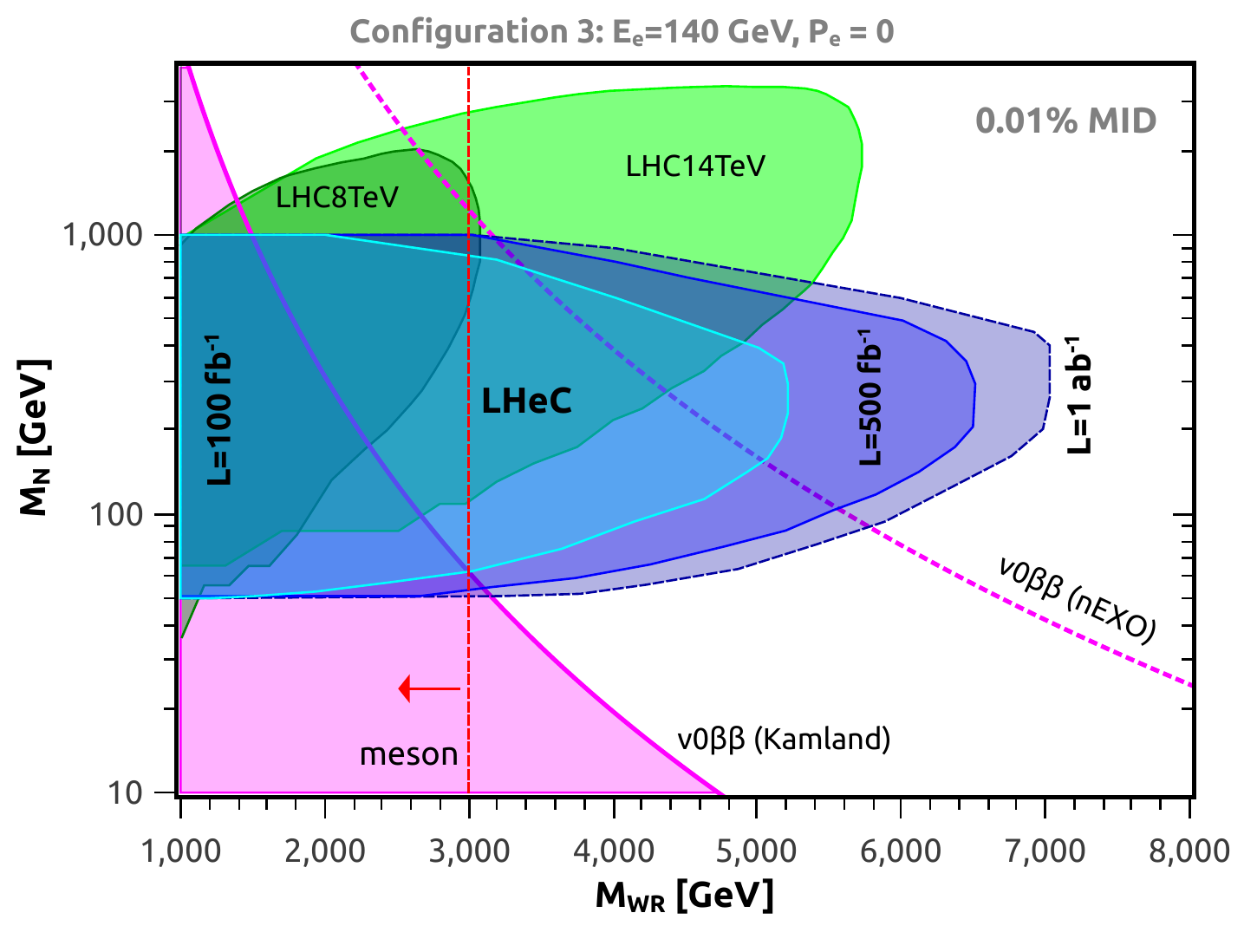}
\caption{{\bf Configuration 3}: 95\% C.L.\ region probed by LHeC for $\mathcal{L}=100$ fb$^{-1}$ (light blue), $500$ fb$^{-1}$ (blue), $1000$ fb$^{-1}$ (dark blue). {\it Left:} assumes $1\%$ MID rate; {\it Right:} adopts a 0.1\% MID rate; {\it Bottom:} uses a 0.01\% MID rate. We have superimposed current limits from $eejj$ (dark green), KamLAND-Zen measurement (shaded pink); and projected limits from: $eejj$ LHC data (light green) and nEXO (dashed pink line) sensitivity.}
\label{140GeVPe00}
\end{figure*}

Let us now examine the impact of increasing the energy of the electron beam to 140 GeV for unpolarized electrons -- referred to as configuration 3. Comparing to configuration 1, we find that the number of signal events increases by a factor of two up to eight depending on the right-handed neutrino mass, whereas the number of background events passing our selection cuts triples. One could possibly impose harder cuts for configuration 3 (also for configuration 4) to reduce even further the SM background, but that would also dwindle the signal strength, which is not so large. Since we are enforcing the presence of at least one signal event at 95\% C.L.\ the LHeC reach would be degraded. 

The LHeC sensitivity regions for this configuration are shown in  Fig.\ \ref{140GeVPe00}, following the same conventions as before.  These regions extend, for 10 years of data, up to  $W_R$ masses of about $4$, $5.5$ and $7$ TeV for a MID rate respectively of $1\%$, $0.1\%$ and $0.01\%$. If the MID rate is $0.1\%$ (right panel), the LHeC in this configuration would be able to probe regions beyond the expected reach of the LHC after the first year, and after 5 years would be exploring areas that can neither be reached by nEXO. This discovery region features Majorana masses between  $100$ and $300$ GeV, and $W_R$ masses between  $4$ and $5.5$ TeV.  Notice also that the $W_R$ mass reach of the LHeC extends beyond that expected at the LHC only for the optimistic MID rate (bottom panel).

If we now compare, for each MID rate, the sensitivity regions with those we obtained in the previous subsection for the  configuration 2 (figure \ref{60GeVPe04}), it becomes evident that the configuration 3 can probe larger regions of the parameter space. That is, if we had to choose between configuration 2 ($E_e=60$ GeV, $P_e=80\%$) and configuration 3 ($E_e=140$ GeV, unpolarized), the latter would be a better choice -- at least regarding the prospects  of probing new regions of parameter space of the left-right model.


\subsection{Configuration 4: $E_e =140$~GeV, $P_e= 80\%$}
\begin{figure*}[!t]
\includegraphics[width=0.5\columnwidth]{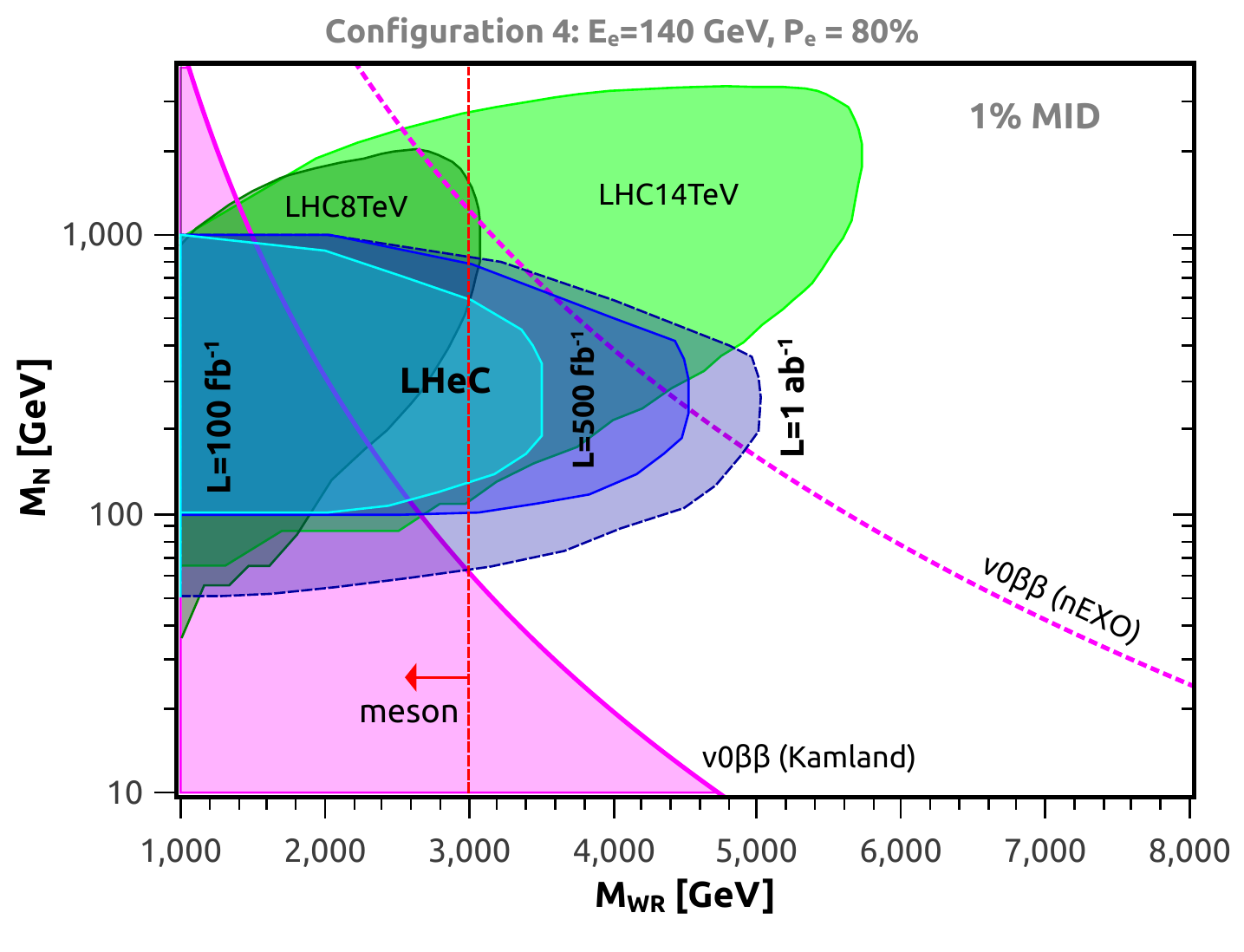}
\includegraphics[width=0.5\columnwidth]{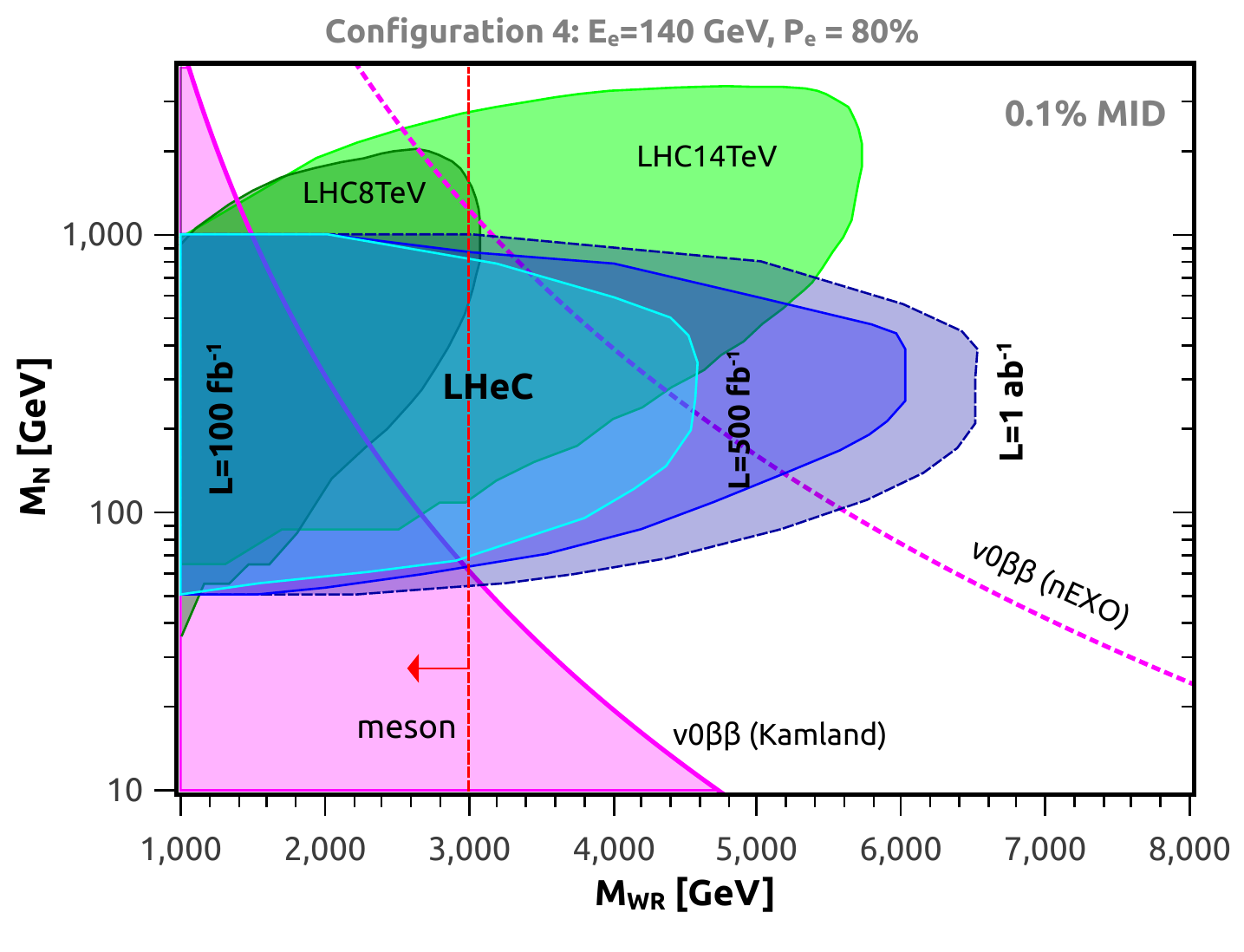}
\center
\includegraphics[width=0.5\columnwidth]{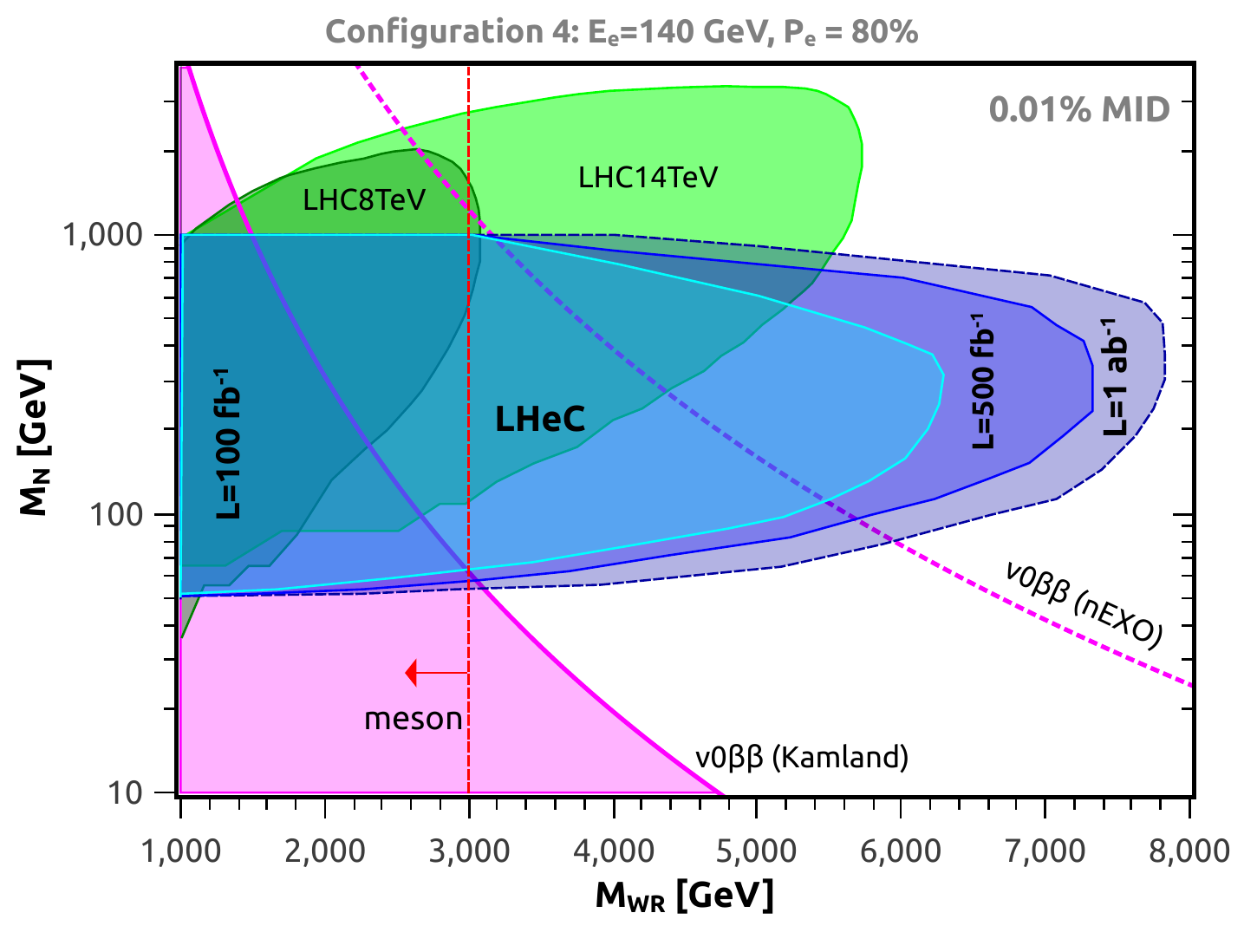}
\caption{{\bf Configuration 4}: 95\% C.L.\ region probed by LHeC for $\mathcal{L}=100$ fb$^{-1}$ (light blue), $500$ fb$^{-1}$ (blue), $1$ ab$^{-1}$  (dark blue). {\it Left:} assumes $1\%$ MID rate; {\it Right:} adopts a 0.1\% MID rate; {\it Bottom:} uses a 0.01\% MID rate. We have superimposed current limits from $eejj$ (dark green), Kamland-Zen measurement (shaded pink); and projected limits from: $eejj$ LHC data (light green) and nEXO (dashed pink line) sensitivity.}
\label{140GeVPe04}
\end{figure*}

Let us now examine the most optimistic configuration, keeping the electron energy at 140 GeV but assuming that  the electron beam is $80\%$ polarized. The effect is similar to configuration 2. It enhances the signal cross section while diminishing the SM background.

Figure \ref{140GeVPe04} shows the resulting  sensitivity regions following our conventions.  Thanks to  the electron polarization, the  LHeC in this configuration can probe heavier $W_R$ gauge bosons and heavier Majorana neutrinos.

For this configuration, the LHeC may start probing regions of the parameter space beyond the expected reach of future experiments even for a conservative MID rate of $1\%$ -- see left panel. Most of that signal region would in any case be within the sensitivity of the LHC and nEXO, so the LHeC would mostly play a complementary role for such a MID rate. But if the MID rate goes down to $0.01\%$ (bottom panel), an admittedly optimistic value,  the reach of the LHec would extend  to $W_R$ masses of almost $8$ TeV, well beyond the reach of other experiments. In this case, the LHeC would be expected to play a leading role in the discovery of the left-right symmetry, probing new and sizable regions of the parameter space with just one year of data taking. For the more realistic MID rate of $0.1\%$,  regions inaccessible to current and planned searches  can be explored after 5 years of data, reaching $W_R$ masses of up to $6.5$ TeV after 10 years.

  
\section{Discussion}

Our results demonstrate the potential of the proposed LHeC collider as far as probing TeV realizations of the left-right symmetric model. These findings rely on some assumptions and are therefore subject to few caveats:

(i) We have assumed for the double beta decay limits that the standard light neutrino mass mechanism is subleading, but its presence could actually 
cancel partly or fully the right-handed Majorana contribution.  
Moreover, there are other diagrams for double beta decay, which can also be related to LHC and LHeC processes. They would introduce, however, additional parameters,  making the study more lengthy and not necessarily more illuminating. 

(ii) Our study was focused on the minimal left-right model. However, other interesting realizations exist which are beyond the scope of this work. We stress though that our reasoning relies on the right-handed current, which is common to all left-right models.

(iii) We have compared, throughout this work, the LHeC sensitivity against the  LHC 14 TeV and nEXO planned sensitivities. It is possible, though, that the actual sensitivities of these experiments turn out to be different, modifying the potential role of the LHeC -- for better or worse.  

(iv) We performed our analysis assuming a 100\% flat rate for the acceptance times efficiency for both signal and background. Thus, our results may be a bit optimistic. In the near future with the release of detailed detector response for such events a more realistic analysis can be done.

With these caveats in mind,  we draw our conclusions below.


\section{Conclusions}
\label{sec:con}
We have investigated the  potential sensitivity of the Large Hadron electron Collider (LHeC) to lepton number violation in the context of minimal left-right symmetric models. The  LHeC is a proposed electron-proton collider that will combine the LHC proton beam of $7$ TeV energy with a $60-140$~GeV electron beam  that can be up to $80\%$ polarized.  In our analysis, we considered  four possible LHeC configurations encompassing different electron energies ($E_e=60$~GeV and $E_e=140$~GeV) and polarizations ($P_e=0$ and $P_e= 80\%$). Our results consists of the regions, in the plane ($M_{W_R}$, $M_{N_R}$), that can be probed at $95\%$ 
C.L.\ with one, five, or ten years of data for each LHeC configuration.

The signal we studied  is the lepton number violating process $e^- p \to e^+ \,3j$, which is the analogue of neutrinoless double beta decay and the LHC process $pp \to 2e \, 2j$. Due to charge misidentification,  this signal might  still suffer from a large detector background in the form of the process $e^- p \to e^- \,3j$. Thus, the charge misidentification (MID) rate at the LHeC played a crucial role in the entire analysis. In our figures, we displayed the sensitivity regions for three different values of this rate, namely  $1\%$, $0.1\%$ and $0.01\%$. 

To determine the possible role of the LHeC in the search for the left-right symmetry and lepton number violation, we compared our sensitivity regions against the current and expected ones from  the LHC and neutrinoless double beta decay experiments. Interestingly, we find that, depending on the configuration and MID rate, the LHeC could cover a sizable region of the parameter space inaccessible to other searches. In addition, the LHeC may also play an important complementary role by confirming a new physics signal previously observed in other experiments, and establishing its possible left-right origin. To do so, a polarized electron beam at the LHeC would be  crucial.

Summarizing, our results show that LHeC is indeed an attractive machine in the quest for left-right symmetry and lepton number violation.  

\acknowledgments
Work supported by the DFG in the Heisenberg programme with grant RO 2516/6-1 (WR), 
and by the Max Planck Society in the project MANITOP (WR, CY).  We warmly thank Bhupal Dev for several discussions as well as Iris Abt, Peter Kostka, Paul Newman for correspondence regarding charge flip misidentification rate.

\bibliographystyle{JHEP}
\bibliography{darkmatter}

\providecommand{\href}[2]{#2}\begingroup\raggedright\begin{thebibliography}{10}

\bibitem{Pati:1974yy}
J.~C. Pati and A.~Salam, {\it {Lepton Number as the Fourth Color}},  {\em Phys.
  Rev.} {\bf D10} (1974) 275--289. [Erratum: Phys. Rev.D11,703(1975)].

\bibitem{Mohapatra:1974hk}
R.~N. Mohapatra and J.~C. Pati, {\it {Left-Right Gauge Symmetry and an
  Isoconjugate Model of CP Violation}},  {\em Phys. Rev.} {\bf D11} (1975)
  566--571.

\bibitem{Mohapatra:1974gc}
R.~N. Mohapatra and J.~C. Pati, {\it {A Natural Left-Right Symmetry}},  {\em
  Phys. Rev.} {\bf D11} (1975) 2558.

\bibitem{Senjanovic:1975rk}
G.~Senjanovic and R.~N. Mohapatra, {\it {Exact Left-Right Symmetry and
  Spontaneous Violation of Parity}},  {\em Phys. Rev.} {\bf D12} (1975) 1502.

\bibitem{Senjanovic:1978ev}
G.~Senjanovic, {\it {Spontaneous Breakdown of Parity in a Class of Gauge
  Theories}},  {\em Nucl. Phys.} {\bf B153} (1979) 334--364.

\bibitem{Minkowski:1977sc}
P.~Minkowski, {\it {$\mu \to e\gamma$ at a Rate of One Out of $10^{9}$ Muon
  Decays?}},  {\em Phys. Lett.} {\bf B67} (1977) 421--428.

\bibitem{Mohapatra:1979ia}
R.~N. Mohapatra and G.~Senjanovic, {\it {Neutrino Mass and Spontaneous Parity
  Violation}},  {\em Phys. Rev. Lett.} {\bf 44} (1980) 912.

\bibitem{Lazarides:1980nt}
G.~Lazarides, Q.~Shafi, and C.~Wetterich, {\it {Proton Lifetime and Fermion
  Masses in an SO(10) Model}},  {\em Nucl. Phys.} {\bf B181} (1981) 287--300.

\bibitem{Mohapatra:1980yp}
R.~N. Mohapatra and G.~Senjanovic, {\it {Neutrino Masses and Mixings in Gauge
  Models with Spontaneous Parity Violation}},  {\em Phys. Rev.} {\bf D23}
  (1981) 165.

\bibitem{Schechter:1980gr}
J.~Schechter and J.~W.~F. Valle, {\it {Neutrino Masses in SU(2) x U(1)
  Theories}},  {\em Phys. Rev.} {\bf D22} (1980) 2227.

\bibitem{Fukugita:1986hr}
M.~Fukugita and T.~Yanagida, {\it {Baryogenesis Without Grand Unification}},
  {\em Phys.Lett.} {\bf B174} (1986) 45.

\bibitem{Berlin:2016eem}
A.~Berlin, P.~J. Fox, D.~Hooper, and G.~Mohlabeng, {\it {Mixed Dark Matter in
  Left-Right Symmetric Models}},  \href{http://xxx.lanl.gov/abs/1604.0610}{{\tt
  arXiv:1604.0610}}.

\bibitem{Borah:2016uoi}
D.~Borah, S.~Patra, and S.~Sahoo, {\it {Subdominant Left-Right Scalar Dark
  Matter as Origin of the 750 GeV Di-photon Excess at LHC}},
  \href{http://xxx.lanl.gov/abs/1601.0182}{{\tt arXiv:1601.0182}}.

\bibitem{Berlin:2016hqw}
A.~Berlin, {\it {Diphoton and diboson excesses in a left-right symmetric theory
  of dark matter}},  {\em Phys. Rev.} {\bf D93} (2016), no.~5 055015,
  [\href{http://xxx.lanl.gov/abs/1601.0138}{{\tt arXiv:1601.0138}}].

\bibitem{Patra:2015vmp}
S.~Patra and S.~Rao, {\it {Singlet fermion Dark Matter within Left-Right
  Model}},  \href{http://xxx.lanl.gov/abs/1512.0405}{{\tt arXiv:1512.0405}}.

\bibitem{Garcia-Cely:2015quu}
C.~Garcia-Cely and J.~Heeck, {\it {Phenomenology of left-right symmetric dark
  matter}},  \href{http://xxx.lanl.gov/abs/1512.0333}{{\tt arXiv:1512.0333}}.

\bibitem{Heeck:2015qra}
J.~Heeck and S.~Patra, {\it {Minimal Left-Right Symmetric Dark Matter}},  {\em
  Phys. Rev. Lett.} {\bf 115} (2015), no.~12 121804,
  [\href{http://xxx.lanl.gov/abs/1507.0158}{{\tt arXiv:1507.0158}}].

\bibitem{Keung:1983uu}
W.-Y. Keung and G.~Senjanovic, {\it {Majorana Neutrinos and the Production of
  the Right-handed Charged Gauge Boson}},  {\em Phys. Rev. Lett.} {\bf 50}
  (1983) 1427.

\bibitem{Chiappetta:1993jy}
P.~Chiappetta, A.~Deliyannis, A.~Fiandrino, and P.~Taxil, {\it {Probing
  left-right symmetric models from right-handed W production with polarized
  beams at LHC}},  {\em Phys. Lett.} {\bf B308} (1993) 304--310.

\bibitem{Maiezza:2010ic}
A.~Maiezza, M.~Nemevsek, F.~Nesti, and G.~Senjanovic, {\it {Left-Right Symmetry
  at LHC}},  {\em Phys. Rev.} {\bf D82} (2010) 055022,
  [\href{http://xxx.lanl.gov/abs/1005.5160}{{\tt arXiv:1005.5160}}].

\bibitem{Tello:2010am}
V.~Tello, M.~Nemevsek, F.~Nesti, G.~Senjanovic, and F.~Vissani, {\it
  {Left-Right Symmetry: from LHC to Neutrinoless Double Beta Decay}},  {\em
  Phys. Rev. Lett.} {\bf 106} (2011) 151801,
  [\href{http://xxx.lanl.gov/abs/1011.3522}{{\tt arXiv:1011.3522}}].

\bibitem{Nemevsek:2011hz}
M.~Nemevsek, F.~Nesti, G.~Senjanovic, and Y.~Zhang, {\it {First Limits on
  Left-Right Symmetry Scale from LHC Data}},  {\em Phys. Rev.} {\bf D83} (2011)
  115014, [\href{http://xxx.lanl.gov/abs/1103.1627}{{\tt arXiv:1103.1627}}].

\bibitem{Esteves:2011gk}
J.~N. Esteves, J.~C. Romao, M.~Hirsch, W.~Porod, F.~Staub, and A.~Vicente, {\it
  {Dark matter and LHC phenomenology in a left-right supersymmetric model}},
  {\em JHEP} {\bf 01} (2012) 095,
  [\href{http://xxx.lanl.gov/abs/1109.6478}{{\tt arXiv:1109.6478}}].

\bibitem{Das:2012ii}
S.~P. Das, F.~F. Deppisch, O.~Kittel, and J.~W.~F. Valle, {\it {Heavy Neutrinos
  and Lepton Flavour Violation in Left-Right Symmetric Models at the LHC}},
  {\em Phys. Rev.} {\bf D86} (2012) 055006,
  [\href{http://xxx.lanl.gov/abs/1206.0256}{{\tt arXiv:1206.0256}}].

\bibitem{Chen:2013fna}
C.-Y. Chen, P.~S.~B. Dev, and R.~N. Mohapatra, {\it {Probing Heavy-Light
  Neutrino Mixing in Left-Right Seesaw Models at the LHC}},  {\em Phys. Rev.}
  {\bf D88} (2013) 033014, [\href{http://xxx.lanl.gov/abs/1306.2342}{{\tt
  arXiv:1306.2342}}].

\bibitem{Arbelaez:2013nga}
C.~Arbeláez, M.~Hirsch, M.~Malinský, and J.~C. Romão, {\it {LHC-scale
  left-right symmetry and unification}},  {\em Phys. Rev.} {\bf D89} (2014),
  no.~3 035002, [\href{http://xxx.lanl.gov/abs/1311.3228}{{\tt
  arXiv:1311.3228}}].

\bibitem{Ferrari:2000sp}
A.~Ferrari, J.~Collot, M.-L. Andrieux, B.~Belhorma, P.~de~Saintignon, J.-Y.
  Hostachy, P.~Martin, and M.~Wielers, {\it {Sensitivity study for new gauge
  bosons and right-handed Majorana neutrinos in $p p$ collisions at $s$ =
  14-TeV}},  {\em Phys. Rev.} {\bf D62} (2000) 013001.

\bibitem{AbelleiraFernandez:2012cc}
{\bf LHeC Study Group} Collaboration, J.~L. Abelleira~Fernandez {\em et.~al.},
  {\it {A Large Hadron Electron Collider at CERN: Report on the Physics and
  Design Concepts for Machine and Detector}},  {\em J. Phys.} {\bf G39} (2012)
  075001, [\href{http://xxx.lanl.gov/abs/1206.2913}{{\tt arXiv:1206.2913}}].

\bibitem{Bruening:2013bga}
O.~Bruening and M.~Klein, {\it {The Large Hadron Electron Collider}},  {\em
  Mod. Phys. Lett.} {\bf A28} (2013), no.~16 1330011,
  [\href{http://xxx.lanl.gov/abs/1305.2090}{{\tt arXiv:1305.2090}}].

\bibitem{Blaksley:2011ey}
C.~Blaksley, M.~Blennow, F.~Bonnet, P.~Coloma, and E.~Fernandez-Martinez, {\it
  {Heavy Neutrinos and Lepton Number Violation in lp Colliders}},  {\em Nucl.
  Phys.} {\bf B852} (2011) 353--365,
  [\href{http://xxx.lanl.gov/abs/1105.0308}{{\tt arXiv:1105.0308}}].

\bibitem{Duarte:2014zea}
L.~Duarte, G.~A. González-Sprinberg, and O.~A. Sampayo, {\it {Majorana
  neutrinos production at LHeC in an effective approach}},  {\em Phys. Rev.}
  {\bf D91} (2015), no.~5 053007,
  [\href{http://xxx.lanl.gov/abs/1412.1433}{{\tt arXiv:1412.1433}}].

\bibitem{Mondal:2015zba}
S.~Mondal and S.~K. Rai, {\it {A polarized window for left-right symmetry at
  the Large Hadron-Electron Collider}},
  \href{http://xxx.lanl.gov/abs/1510.0863}{{\tt arXiv:1510.0863}}.

\bibitem{Queiroz:2016qmc}
F.~S. Queiroz, {\it {Comment on “Polarized window for left-right symmetry and
  a right-handed neutrino at the Large Hadron-Electron Collider”}},  {\em
  Phys. Rev.} {\bf D93} (2016), no.~11 118701.

\bibitem{Mondal:2016czu}
S.~Mondal and S.~K. Rai, {\it {Reply to “Comment on ‘Polarized window for
  left-right symmetry and a right-handed neutrino at the Large Hadron-Electron
  Collider’”}},  {\em Phys. Rev.} {\bf D93} (2016), no.~11 118702.

\bibitem{Asakura:2014lma}
{\bf KamLAND-Zen} Collaboration, K.~Asakura {\em et.~al.}, {\it {Results from
  KamLAND-Zen}},  {\em AIP Conf. Proc.} {\bf 1666} (2015) 170003,
  [\href{http://xxx.lanl.gov/abs/1409.0077}{{\tt arXiv:1409.0077}}].

\bibitem{Pocar:2015mrz}
{\bf nEXO, EXO-200} Collaboration, A.~Pocar, {\it {From EXO-200 to nEXO}},
  {\em PoS} {\bf NEUTEL2015} (2015) 049.

\bibitem{Bambhaniya:2015wna}
G.~Bambhaniya, J.~Chakrabortty, J.~Gluza, T.~Jelinski, and R.~Szafron, {\it
  {Search for doubly charged Higgs bosons through vector boson fusion at the
  LHC and beyond}},  {\em Phys. Rev.} {\bf D92} (2015), no.~1 015016,
  [\href{http://xxx.lanl.gov/abs/1504.0399}{{\tt arXiv:1504.0399}}].

\bibitem{Bambhaniya:2014cia}
G.~Bambhaniya, J.~Chakrabortty, J.~Gluza, T.~Jeliński, and M.~Kordiaczynska,
  {\it {Lowest limits on the doubly charged Higgs boson masses in the minimal
  left-right symmetric model}},  {\em Phys. Rev.} {\bf D90} (2014), no.~9
  095003, [\href{http://xxx.lanl.gov/abs/1408.0774}{{\tt arXiv:1408.0774}}].

\bibitem{Bambhaniya:2013wza}
G.~Bambhaniya, J.~Chakrabortty, J.~Gluza, M.~Kordiaczyńska, and R.~Szafron,
  {\it {Left-Right Symmetry and the Charged Higgs Bosons at the LHC}},  {\em
  JHEP} {\bf 05} (2014) 033, [\href{http://xxx.lanl.gov/abs/1311.4144}{{\tt
  arXiv:1311.4144}}].

\bibitem{Dev:2016dja}
P.~S.~B. Dev, R.~N. Mohapatra, and Y.~Zhang, {\it {Probing the Higgs Sector of
  the Minimal Left-Right Symmetric Model at Future Hadron Colliders}},
  \href{http://xxx.lanl.gov/abs/1602.0594}{{\tt arXiv:1602.0594}}.

\bibitem{Bambhaniya:2015ipg}
G.~Bambhaniya, P.~S.~B. Dev, S.~Goswami, and M.~Mitra, {\it {The Scalar Triplet
  Contribution to Lepton Flavour Violation and Neutrinoless Double Beta Decay
  in Left-Right Symmetric Model}},
  \href{http://xxx.lanl.gov/abs/1512.0044}{{\tt arXiv:1512.0044}}.

\bibitem{Mohapatra:2013cia}
R.~N. Mohapatra and Y.~Zhang, {\it {LHC accessible second Higgs boson in the
  left-right model}},  {\em Phys. Rev.} {\bf D89} (2014), no.~5 055001,
  [\href{http://xxx.lanl.gov/abs/1401.0018}{{\tt arXiv:1401.0018}}].

\bibitem{Patra:2015bga}
S.~Patra, F.~S. Queiroz, and W.~Rodejohann, {\it {Stringent Dilepton Bounds on
  Left-Right Models using LHC data}},  {\em Phys. Lett.} {\bf B752} (2016)
  186--190, [\href{http://xxx.lanl.gov/abs/1506.0345}{{\tt arXiv:1506.0345}}].

\bibitem{Deppisch:2015qwa}
F.~F. Deppisch, P.~S. Bhupal~Dev, and A.~Pilaftsis, {\it {Neutrinos and
  Collider Physics}},  {\em New J. Phys.} {\bf 17} (2015), no.~7 075019,
  [\href{http://xxx.lanl.gov/abs/1502.0654}{{\tt arXiv:1502.0654}}].

\bibitem{Aad:2014cka}
{\bf ATLAS} Collaboration, G.~Aad {\em et.~al.}, {\it {Search for high-mass
  dilepton resonances in pp collisions at $\sqrt{s}=8$  TeV with the ATLAS
  detector}},  {\em Phys. Rev.} {\bf D90} (2014), no.~5 052005,
  [\href{http://xxx.lanl.gov/abs/1405.4123}{{\tt arXiv:1405.4123}}].

\bibitem{Khachatryan:2014dka}
{\bf CMS} Collaboration, V.~Khachatryan {\em et.~al.}, {\it {Search for heavy
  neutrinos and $\mathrm {W}$ bosons with right-handed couplings in
  proton-proton collisions at $\sqrt{s} = 8\,\text {TeV} $}},  {\em Eur. Phys.
  J.} {\bf C74} (2014), no.~11 3149,
  [\href{http://xxx.lanl.gov/abs/1407.3683}{{\tt arXiv:1407.3683}}].

\bibitem{Aad:2015xaa}
{\bf ATLAS} Collaboration, G.~Aad {\em et.~al.}, {\it {Search for heavy
  Majorana neutrinos with the ATLAS detector in pp collisions at $ \sqrt{s}=8 $
  TeV}},  {\em JHEP} {\bf 07} (2015) 162,
  [\href{http://xxx.lanl.gov/abs/1506.0602}{{\tt arXiv:1506.0602}}].

\bibitem{Buchmuller:1991tu}
W.~Buchmuller and C.~Greub, {\it {Heavy Majorana neutrinos in electron -
  positron and electron - proton collisions}},  {\em Nucl. Phys.} {\bf B363}
  (1991) 345--368.

\bibitem{Ingelman:1993ve}
G.~Ingelman and J.~Rathsman, {\it {Heavy Majorana neutrinos at e p colliders}},
   {\em Z. Phys.} {\bf C60} (1993) 243--254.

\bibitem{Buchmuller:1990vh}
W.~Buchmuller and C.~Greub, {\it {Electroproduction of Majorana neutrinos}},
  {\em Phys. Lett.} {\bf B256} (1991) 465--470.

\bibitem{Buchmuller:1992wm}
W.~Buchmuller and C.~Greub, {\it {Right-handed currents and heavy neutrinos in
  high-energy $e p$ and $e^{+} e^{-}$ scattering}},  {\em Nucl. Phys.} {\bf
  B381} (1992) 109--128.

\bibitem{Flanz:1999ah}
M.~Flanz, W.~Rodejohann, and K.~Zuber, {\it {Bounds on effective Majorana
  neutrino masses at HERA}},  {\em Phys. Lett.} {\bf B473} (2000) 324--329,
  [\href{http://xxx.lanl.gov/abs/hep-ph/9911298}{{\tt hep-ph/9911298}}].
  [Erratum: Phys. Lett.B480,418(2000)].

\bibitem{Helo:2013ika}
J.~C. Helo, M.~Hirsch, H.~Päs, and S.~G. Kovalenko, {\it {Short-range
  mechanisms of neutrinoless double beta decay at the LHC}},  {\em Phys. Rev.}
  {\bf D88} (2013) 073011, [\href{http://xxx.lanl.gov/abs/1307.4849}{{\tt
  arXiv:1307.4849}}].

\bibitem{Maiezza:2014ala}
A.~Maiezza and M.~Nemevšek, {\it {Strong P invariance, neutron electric dipole
  moment, and minimal left-right parity at LHC}},  {\em Phys. Rev.} {\bf D90}
  (2014), no.~9 095002, [\href{http://xxx.lanl.gov/abs/1407.3678}{{\tt
  arXiv:1407.3678}}].

\bibitem{Fowlie:2014mza}
A.~Fowlie and L.~Marzola, {\it {Testing quark mixing in minimal left–right
  symmetric models with b -tags at the LHC}},  {\em Nucl. Phys.} {\bf B889}
  (2014) 36--45, [\href{http://xxx.lanl.gov/abs/1408.6699}{{\tt
  arXiv:1408.6699}}].

\bibitem{Dutta:2014dba}
B.~Dutta, R.~Eusebi, Y.~Gao, T.~Ghosh, and T.~Kamon, {\it {Exploring the doubly
  charged Higgs boson of the left-right symmetric model using vector boson
  fusionlike events at the LHC}},  {\em Phys. Rev.} {\bf D90} (2014) 055015,
  [\href{http://xxx.lanl.gov/abs/1404.0685}{{\tt arXiv:1404.0685}}].

\bibitem{Parida:2014dla}
M.~K. Parida and B.~Sahoo, {\it {Planck-scale induced left–right gauge theory
  at LHC and experimental tests}},  {\em Nucl. Phys.} {\bf B906} (2016)
  77--104, [\href{http://xxx.lanl.gov/abs/1411.6748}{{\tt arXiv:1411.6748}}].

\bibitem{Helo:2015ffa}
J.~C. Helo and M.~Hirsch, {\it {LHC dijet constraints on double beta decay}},
  {\em Phys. Rev.} {\bf D92} (2015), no.~7 073017,
  [\href{http://xxx.lanl.gov/abs/1509.0042}{{\tt arXiv:1509.0042}}].

\bibitem{Gluza:2015goa}
J.~Gluza and T.~Jeliński, {\it {Heavy neutrinos and the pp→lljj CMS data}},
  {\em Phys. Lett.} {\bf B748} (2015) 125--131,
  [\href{http://xxx.lanl.gov/abs/1504.0556}{{\tt arXiv:1504.0556}}].

\bibitem{Brehmer:2015cia}
J.~Brehmer, J.~Hewett, J.~Kopp, T.~Rizzo, and J.~Tattersall, {\it {Symmetry
  Restored in Dibosons at the LHC?}},  {\em JHEP} {\bf 10} (2015) 182,
  [\href{http://xxx.lanl.gov/abs/1507.0001}{{\tt arXiv:1507.0001}}].

\bibitem{Deppisch:2015cua}
F.~F. Deppisch, L.~Graf, S.~Kulkarni, S.~Patra, W.~Rodejohann, N.~Sahu, and
  U.~Sarkar, {\it {Reconciling the 2 TeV excesses at the LHC in a linear seesaw
  left-right model}},  {\em Phys. Rev.} {\bf D93} (2016), no.~1 013011,
  [\href{http://xxx.lanl.gov/abs/1508.0594}{{\tt arXiv:1508.0594}}].

\bibitem{Gluza:2016qqv}
J.~Gluza, T.~Jelinski, and R.~Szafron, {\it {Lepton Number Violation and
  `Diracness' of massive neutrinos composed of Majorana states}},
  \href{http://xxx.lanl.gov/abs/1604.0138}{{\tt arXiv:1604.0138}}.

\bibitem{Lindner:2016lpp}
M.~Lindner, F.~S. Queiroz, and W.~Rodejohann, {\it {Dilepton bounds on
  left-right symmetry at the LHC run II and neutrinoless double beta decay}},
  \href{http://xxx.lanl.gov/abs/1604.0741}{{\tt arXiv:1604.0741}}.

\bibitem{Vergados:2002pv}
J.~D. Vergados, {\it {The Neutrinoless double beta decay from a modern
  perspective}},  {\em Phys. Rept.} {\bf 361} (2002) 1--56,
  [\href{http://xxx.lanl.gov/abs/hep-ph/0209347}{{\tt hep-ph/0209347}}].

\bibitem{Simkovic:2007vu}
F.~Simkovic, A.~Faessler, V.~Rodin, P.~Vogel, and J.~Engel, {\it {Anatomy of
  nuclear matrix elements for neutrinoless double-beta decay}},  {\em Phys.
  Rev.} {\bf C77} (2008) 045503, [\href{http://xxx.lanl.gov/abs/0710.2055}{{\tt
  arXiv:0710.2055}}].

\bibitem{Avignone:2007fu}
F.~T. Avignone, III, S.~R. Elliott, and J.~Engel, {\it {Double Beta Decay,
  Majorana Neutrinos, and Neutrino Mass}},  {\em Rev. Mod. Phys.} {\bf 80}
  (2008) 481--516, [\href{http://xxx.lanl.gov/abs/0708.1033}{{\tt
  arXiv:0708.1033}}].

\bibitem{Rodejohann:2011mu}
W.~Rodejohann, {\it {Neutrino-less Double Beta Decay and Particle Physics}},
  {\em Int. J. Mod. Phys.} {\bf E20} (2011) 1833--1930,
  [\href{http://xxx.lanl.gov/abs/1106.1334}{{\tt arXiv:1106.1334}}].

\bibitem{Elliott:2012sp}
S.~R. Elliott, {\it {Recent Progress in Double Beta Decay}},  {\em Mod. Phys.
  Lett.} {\bf A27} (2012) 1230009,
  [\href{http://xxx.lanl.gov/abs/1203.1070}{{\tt arXiv:1203.1070}}].

\bibitem{Bilenky:2012qi}
S.~M. Bilenky and C.~Giunti, {\it {Neutrinoless double-beta decay: A brief
  review}},  {\em Mod. Phys. Lett.} {\bf A27} (2012) 1230015,
  [\href{http://xxx.lanl.gov/abs/1203.5250}{{\tt arXiv:1203.5250}}].

\bibitem{Vergados:2012xy}
J.~D. Vergados, H.~Ejiri, and F.~Simkovic, {\it {Theory of Neutrinoless Double
  Beta Decay}},  {\em Rept. Prog. Phys.} {\bf 75} (2012) 106301,
  [\href{http://xxx.lanl.gov/abs/1205.0649}{{\tt arXiv:1205.0649}}].

\bibitem{Rodejohann:2012xd}
W.~Rodejohann, {\it {Neutrinoless double beta decay and neutrino physics}},
  {\em J. Phys.} {\bf G39} (2012) 124008,
  [\href{http://xxx.lanl.gov/abs/1206.2560}{{\tt arXiv:1206.2560}}].

\bibitem{Deppisch:2012nb}
F.~F. Deppisch, M.~Hirsch, and H.~Pas, {\it {Neutrinoless Double Beta Decay and
  Physics Beyond the Standard Model}},  {\em J. Phys.} {\bf G39} (2012) 124007,
  [\href{http://xxx.lanl.gov/abs/1208.0727}{{\tt arXiv:1208.0727}}].

\bibitem{Vogel:2012ja}
P.~Vogel, {\it {Nuclear structure and double beta decay}},  {\em J. Phys.} {\bf
  G39} (2012) 124002, [\href{http://xxx.lanl.gov/abs/1208.1992}{{\tt
  arXiv:1208.1992}}].

\bibitem{Schwingenheuer:2012zs}
B.~Schwingenheuer, {\it {Status and prospects of searches for neutrinoless
  double beta decay}},  {\em Annalen Phys.} {\bf 525} (2013) 269--280,
  [\href{http://xxx.lanl.gov/abs/1210.7432}{{\tt arXiv:1210.7432}}].

\bibitem{Petcov:2013poa}
S.~T. Petcov, {\it {The Nature of Massive Neutrinos}},  {\em Adv. High Energy
  Phys.} {\bf 2013} (2013) 852987,
  [\href{http://xxx.lanl.gov/abs/1303.5819}{{\tt arXiv:1303.5819}}].

\bibitem{Cremonesi:2013vla}
O.~Cremonesi and M.~Pavan, {\it {Challenges in Double Beta Decay}},  {\em Adv.
  High Energy Phys.} {\bf 2014} (2014) 951432,
  [\href{http://xxx.lanl.gov/abs/1310.4692}{{\tt arXiv:1310.4692}}].

\bibitem{Pas:2015eia}
H.~Pas and W.~Rodejohann, {\it {Neutrinoless Double Beta Decay}},  {\em New J.
  Phys.} {\bf 17} (2015), no.~11 115010,
  [\href{http://xxx.lanl.gov/abs/1507.0017}{{\tt arXiv:1507.0017}}].

\bibitem{KlapdorKleingrothaus:2000sn}
H.~V. Klapdor-Kleingrothaus {\em et.~al.}, {\it {Latest results from the
  Heidelberg-Moscow double beta decay experiment}},  {\em Eur. Phys. J.} {\bf
  A12} (2001) 147--154, [\href{http://xxx.lanl.gov/abs/hep-ph/0103062}{{\tt
  hep-ph/0103062}}].

\bibitem{Gando:2012zm}
{\bf KamLAND-Zen} Collaboration, A.~Gando {\em et.~al.}, {\it {Limit on
  Neutrinoless $\beta\beta$ Decay of $^{136}$Xe from the First Phase of
  KamLAND-Zen and Comparison with the Positive Claim in $^{76}$Ge}},  {\em
  Phys. Rev. Lett.} {\bf 110} (2013), no.~6 062502,
  [\href{http://xxx.lanl.gov/abs/1211.3863}{{\tt arXiv:1211.3863}}].

\bibitem{Auger:2012ar}
{\bf EXO-200} Collaboration, M.~Auger {\em et.~al.}, {\it {Search for
  Neutrinoless Double-Beta Decay in $^{136}$Xe with EXO-200}},  {\em Phys. Rev.
  Lett.} {\bf 109} (2012) 032505,
  [\href{http://xxx.lanl.gov/abs/1205.5608}{{\tt arXiv:1205.5608}}].

\bibitem{Agostini:2013mzu}
{\bf GERDA} Collaboration, M.~Agostini {\em et.~al.}, {\it {Results on
  Neutrinoless Double-$\beta$ Decay of $^{76}$Ge from Phase I of the GERDA
  Experiment}},  {\em Phys. Rev. Lett.} {\bf 111} (2013), no.~12 122503,
  [\href{http://xxx.lanl.gov/abs/1307.4720}{{\tt arXiv:1307.4720}}].

\bibitem{Albert:2014awa}
{\bf EXO-200} Collaboration, J.~B. Albert {\em et.~al.}, {\it {Search for
  Majorana neutrinos with the first two years of EXO-200 data}},  {\em Nature}
  {\bf 510} (2014) 229--234, [\href{http://xxx.lanl.gov/abs/1402.6956}{{\tt
  arXiv:1402.6956}}].

\bibitem{Dev:2013vxa}
P.~S. Bhupal~Dev, S.~Goswami, M.~Mitra, and W.~Rodejohann, {\it {Constraining
  Neutrino Mass from Neutrinoless Double Beta Decay}},  {\em Phys. Rev.} {\bf
  D88} (2013) 091301, [\href{http://xxx.lanl.gov/abs/1305.0056}{{\tt
  arXiv:1305.0056}}].

\bibitem{Dev:2014xea}
P.~Bhupal~Dev, S.~Goswami, and M.~Mitra, {\it {TeV Scale Left-Right Symmetry
  and Large Mixing Effects in Neutrinoless Double Beta Decay}},  {\em Phys.
  Rev.} {\bf D91} (2015), no.~11 113004,
  [\href{http://xxx.lanl.gov/abs/1405.1399}{{\tt arXiv:1405.1399}}].

\bibitem{Dev:2015vra}
P.~S. Bhupal~Dev, C.-H. Lee, and R.~N. Mohapatra, {\it {TeV Scale Lepton Number
  Violation and Baryogenesis}},  {\em J. Phys. Conf. Ser.} {\bf 631} (2015),
  no.~1 012007, [\href{http://xxx.lanl.gov/abs/1503.0497}{{\tt
  arXiv:1503.0497}}].

\bibitem{Awasthi:2015ota}
R.~L. Awasthi, P.~S.~B. Dev, and M.~Mitra, {\it {Implications of the Diboson
  Excess for Neutrinoless Double Beta Decay and Lepton Flavor Violation in TeV
  Scale Left Right Symmetric Model}},  {\em Phys. Rev.} {\bf D93} (2016), no.~1
  011701, [\href{http://xxx.lanl.gov/abs/1509.0538}{{\tt arXiv:1509.0538}}].

\bibitem{Nemevsek:2011aa}
M.~Nemevsek, F.~Nesti, G.~Senjanovic, and V.~Tello, {\it {Neutrinoless Double
  Beta Decay: Low Left-Right Symmetry Scale?}},
  \href{http://xxx.lanl.gov/abs/1112.3061}{{\tt arXiv:1112.3061}}.

\bibitem{Peng:2015haa}
T.~Peng, M.~J. Ramsey-Musolf, and P.~Winslow, {\it {TeV Lepton Number
  Violation: From Neutrinoless Double Beta Decay to the LHC}},
  \href{http://xxx.lanl.gov/abs/1508.0444}{{\tt arXiv:1508.0444}}.

\bibitem{Han:2012vk}
T.~Han, I.~Lewis, R.~Ruiz, and Z.-g. Si, {\it {Lepton Number Violation and
  $W^\prime$ Chiral Couplings at the LHC}},  {\em Phys. Rev.} {\bf D87} (2013),
  no.~3 035011, [\href{http://xxx.lanl.gov/abs/1211.6447}{{\tt
  arXiv:1211.6447}}]. [Erratum: Phys. Rev.D87,no.3,039906(2013)].

\bibitem{Helo:2013dla}
J.~C. Helo, M.~Hirsch, S.~G. Kovalenko, and H.~Pas, {\it {Neutrinoless double
  beta decay and lepton number violation at the LHC}},  {\em Phys. Rev.} {\bf
  D88} (2013), no.~1 011901, [\href{http://xxx.lanl.gov/abs/1303.0899}{{\tt
  arXiv:1303.0899}}].

\bibitem{Teixeira:2014jza}
A.~M. Teixeira, A.~Abada, A.~J.~R. Figueiredo, and J.~C. Romao, {\it {Lepton
  flavour violation at high energies: the LHC and a Linear Collider}},  {\em
  Nuovo Cim.} {\bf C037} (2014), no.~02 19--24,
  [\href{http://xxx.lanl.gov/abs/1402.1426}{{\tt arXiv:1402.1426}}].

\bibitem{Abada:2013bpa}
A.~Abada, {\it {Neutrino Physics, Lepton Flavour Violation and the LHC}},  in
  {\em {25th Rencontres de Blois on Particle Physics and Cosmology Blois,
  France, May 26-31, 2013}}, 2013.
\newblock \href{http://xxx.lanl.gov/abs/1310.3800}{{\tt arXiv:1310.3800}}.

\bibitem{Abada:2012re}
A.~Abada, A.~J.~R. Figueiredo, J.~C. Romao, and A.~M. Teixeira, {\it {Lepton
  flavour violation: physics potential of a Linear Collider}},  {\em JHEP} {\bf
  08} (2012) 138, [\href{http://xxx.lanl.gov/abs/1206.2306}{{\tt
  arXiv:1206.2306}}].

\bibitem{Abada:2008gs}
A.~Abada, P.~Hosteins, F.-X. Josse-Michaux, and S.~Lavignac, {\it {Successful
  Leptogenesis in SO(10) Unification with a Left-Right Symmetric Seesaw
  Mechanism}},  {\em Nucl. Phys.} {\bf B809} (2009) 183--217,
  [\href{http://xxx.lanl.gov/abs/0808.2058}{{\tt arXiv:0808.2058}}].

\bibitem{Barry:2013xxa}
J.~Barry and W.~Rodejohann, {\it {Lepton number and flavour violation in
  TeV-scale left-right symmetric theories with large left-right mixing}},  {\em
  JHEP} {\bf 09} (2013) 153, [\href{http://xxx.lanl.gov/abs/1303.6324}{{\tt
  arXiv:1303.6324}}].

\bibitem{Khachatryan:2015hwa}
{\bf CMS} Collaboration, V.~Khachatryan {\em et.~al.}, {\it {Performance of
  Electron Reconstruction and Selection with the CMS Detector in Proton-Proton
  Collisions at √s = 8 TeV}},  {\em JINST} {\bf 10} (2015), no.~06 P06005,
  [\href{http://xxx.lanl.gov/abs/1502.0270}{{\tt arXiv:1502.0270}}].

\bibitem{Belyaev:2012qa}
A.~Belyaev, N.~D. Christensen, and A.~Pukhov, {\it {CalcHEP 3.4 for collider
  physics within and beyond the Standard Model}},  {\em Comput. Phys. Commun.}
  {\bf 184} (2013) 1729--1769, [\href{http://xxx.lanl.gov/abs/1207.6082}{{\tt
  arXiv:1207.6082}}].

\bibitem{Kniehl:1990iv}
B.~A. Kniehl, {\it {Elastic e p scattering and the Weizsacker-Williams
  approximation}},  {\em Phys. Lett.} {\bf B254} (1991) 267--273.

\bibitem{Sjostrand:2014zea}
T.~Sjöstrand, S.~Ask, J.~R. Christiansen, R.~Corke, N.~Desai, P.~Ilten,
  S.~Mrenna, S.~Prestel, C.~O. Rasmussen, and P.~Z. Skands, {\it {An
  Introduction to PYTHIA 8.2}},  {\em Comput. Phys. Commun.} {\bf 191} (2015)
  159--177, [\href{http://xxx.lanl.gov/abs/1410.3012}{{\tt arXiv:1410.3012}}].

\end{thebibliography}\endgroup

\end{document}